\documentclass[final,5p,times,twocolumn]{elsarticle}

\usepackage{amssymb}
\usepackage{amsthm}

\usepackage{lineno,hyperref}

\usepackage{xcolor}
\usepackage{booktabs}

\usepackage{threeparttable}

\journal{Vacuum}

\begin{document}

\begin{frontmatter}



\title{Influence of Specific Energy Inhomogeneity on the CO\textsubscript{2} Splitting Performance in a High-Power Plasma Jet}


\author{Hendrik Burghaus}
\author{Clemens F. Kaiser}
\author{Stefanos Fasoulas}
\author{Georg Herdrich}

\affiliation{organization={Institute of Space Systems (IRS), University of Stuttgart},
            addressline={Pfaffenwaldring 29}, 
            city={Stuttgart},
            postcode={70569}, 
            state={Baden-Württemberg},
            country={Germany}}

\begin{abstract}
Plasma-based CO\textsubscript{2} conversion is a promising pathway towards greenhouse gas recycling. In the corresponding research field, various types of plasma reactors are applied for carbon dioxide dissociation. So far, spatial inhomogeneities of the specific energy (SEI) distribution in plasma generators, e.g., induced by non-uniform heating or an inhomogeneous mass distribution, are not the focus of the investigations. In this work, the spatial inhomogeneity of mass-specific enthalpy in the plasma jet of the inductive plasma generator IPG4 at the Institute of Space Systems (IRS) is examined. For this purpose, the mean mass-specific enthalpy as well as the radial distribution of the local enthalpy are measured using plasma probes. Moreover, the influence of the determined specific enthalpy inhomogeneity on the CO\textsubscript{2} splitting performance is quantified. It is shown that an inhomogeneous radial distribution of the specific energy can significantly lower the carbon dioxide conversion, compared to a homogeneous case. With regards to IPG4, the performance reduction is $16\,\%$.

\end{abstract}




\end{frontmatter}


\section{Introduction}
\label{sec:introduction}
The fundamental role of anthropogenic greenhouse gas (GHG) emissions in the relentlessly progressing climate change on Earth is undeniable. Therefore, the Intergovernmental Panel on Climate Change (IPCC) sees a massive reduction in GHG emissions as the only way to limit global warming to $1.5\,^\circ\mathrm{C}$ or even $2.0\,^\circ\mathrm{C}$. In the IPCC's sixth assessment report, Carbon Capture and Utilization (CCU) and Carbon Capture and Storage (CCS) are defined as necessary tools for the mitigation of carbon dioxide emissions \cite{IPCC2022}. To this end, plasma technology might be a promising way of converting CO\textsubscript{2} into value-added compounds \cite{Bogaerts2020}. One fundamental process under investigation is the plasma-based conversion of pure carbon dioxide into the syngas carbon monoxide and oxygen, called CO\textsubscript{2} splitting \cite{Snoeckx2017}:

\begin{equation}
	\mathit{CO_2} \rightarrow \mathit{CO} + \frac{1}{2} \mathit{O_2}  \quad \Delta H^\mathrm{0}_\mathrm{298} = 2.93\,\mathrm{eV/molecule}
\end{equation}

Major advancements have been made in the application of non-thermal plasmas (NTP) for CO\textsubscript{2} splitting, with and without catalyst \cite{George2021}. Besides NTP, also the dissociation of CO\textsubscript{2} in thermodynamic equilibrium is increasingly subject of research, as it could be demonstrated that thermal reactions dominate the CO\textsubscript{2} conversion in microwave (MW) and gliding arc (GA) plasmas \cite{Bogaerts2020}. Here, several ways to overcome the energy efficiency limit of around $50\,\%$ \cite{Snoeckx2017} have been identified, such as super-ideal chemical quenching by O-CO\textsubscript{2} association \cite{Van2021} or thermal quenching by fast expansion through a de Laval nozzle \cite{Vermeiren2018, Hecimovic2022}. \par

A great part of the research is focused on the reaction kinetics, especially the excitation of vibrational modes of carbon dioxide. In this context of plasma-based CO\textsubscript{2} splitting, the specific energy input (SEI) has been identified as a key parameter, which is commonly expressed in units of eV/molecule \cite{Snoeckx2017}:

\begin{equation}
	\label{eq:SEI_definition}
	\mathit{SEI}[\mathrm{eV/molecule}] = \frac{P_\mathrm{cal}}{\dot{m}_\mathrm{CO2}} \cdot \frac{M_\mathrm{CO2}}{e \cdot N_\mathrm{A}}
\end{equation}

Here, $P_\mathrm{cal}$ is the plasma power, $\dot{m}_\mathrm{CO2}$ the mass flow rate of carbon dioxide, $M_\mathrm{CO2}$ is the molar mass of CO\textsubscript{2}, $e$ is the elementary charge and $N_\mathrm{A}$ is the Avogadro's constant. In the shown formulation, Eq. \ref{eq:SEI_definition} is applicable for a pure carbon dioxide gas. While current research concentrates on the optimization of the specific energy input, not much can be found on spatial inhomogeneities of the SEI in a plasma jet in literature. Wolf et al. numerically simulate non-uniform heating in a MW plasma generator and stress the importance of local SEI measurements in contrast to the commonly used global values \cite{Wolf2020}. To date, local effects are barely understood, especially from the experimental side. Nevertheless, the occurrences of SEI inhomogeneities are described in a variety of studies. Possible origins of the inhomogeneities are a spatially non-uniform power distribution in the discharge region, heterogeneous distribution of the mass flux, expansion of the plume, wall cooling effects, magnetohydrodynamic (MHD) effects, or, as is usually the case, the combination of several of these. In general, most types of plasma generators are expected to show spatial SEI inhomogeneities due to a complex superposition of mass flux and power distribution. In this regard, Bogaerts and Centi state that in GA plasmas only a fraction of the gas passes the discharge, which limits the conversion \cite{Bogaerts2020}. Moreover, contraction phenomena are known to be a limiting factor in MW plasmas \cite{Bogaerts2020, Viegas2020} with respect to carbon dioxide conversion. \par

At the Institute of Space Systems (IRS), the plasma wind tunnel PWK3, powered by the inductive plasma generator IPG4, is used for experimental investigations on thermal CO\textsubscript{2} splitting at high powers. Specific energy inhomogeneities in the plasma jet of PWK3 are known and measured for two decades already (e.g., \cite{Herdrich2004, Lohle2009, Marynowski2014, Massuti2018, Burghaus2023}). Moreover, mean specific enthalpies are reported for pure oxygen flows in PWK3 by Herdrich \cite{Herdrich2002}. Only recently with the investigations on CO\textsubscript{2} splitting at IRS (\cite{Burghaus2021, Burghaus2023}), the correlation between radially distributed local specific energy and integral values in PWK3 moved into focus.

It has to be noted that in the field of plasma wind tunnels, the energy per particle in the plasma is labeled as total mass-specific enthalpy $h_\mathrm{tot}$. This measure is the same as the specific energy input, but usually specified in different units:

\begin{equation}
	\label{eq:mean_enthalpy}
	h_\mathrm{tot}[\mathrm{J/kg}] =\mathit{SEI}[\mathrm{eV/molecule}]\frac{e \cdot N_\mathrm{A}}{M_\mathrm{CO2}} = \frac{P_\mathrm{cal}}{\dot{m}_\mathrm{CO2}}
\end{equation}

In accordance with Eq. \ref{eq:SEI_definition}), $P_\mathrm{cal}$ is the plasma power and $\dot{m}_\mathrm{CO2}$ the mass flow rate of carbon dioxide. In the course of this work, the terms specific energy input, specific energy, mass-specific enthalpy and specific enthalpy are used interchangeably. The parameters $\mathit{SEI}$ and $h_\mathrm{tot}$ are applied for the expression of the same quantity in units of $[\mathrm{eV/molecule}]$ and $[\mathrm{J/kg}]$, respectively.\par

In this paper, the influence of spatial (radial) specific energy inhomogeneities in the plasma jet of the inductive plasma generator IPG4 on the carbon dioxide splitting performance is examined by comparing integral measurements of the mean (bulk) enthalpy and locally measured mass-specific enthalpy values. \par

In Section \ref{sec:experimental_setup}, the experimental facility as well as the plasma probes used in the course of this work are introduced. Furthermore, a software tool for thermal carbon dioxide splitting is described. Section \ref{sec:results} contains the measurement results of mean and local enthalpies. Moreover, the influence of specific energy inhomogeneities on the CO\textsubscript{2} splitting performance is investigated by a parameter study.

\section{Experimental setup and tools}
\label{sec:experimental_setup}
At the Institute of Space Systems (IRS) three plasma wind tunnels are operated to experimentally simulate the entry of objects into planetary atmospheres \cite{Loehle2021}. All experiments in the course of this work are conducted in the plasma wind tunnel PWK3, powered by the inductively coupled plasma generator IPG4. In the following, the experimental setup of PWK3 and the applied intrusive plasma diagnostics are described. Moreover, the concept of carbon dioxide splitting in thermodynamic equilibrium is introduced.

\subsection{Experimental facility}
\label{sec:experimental_facility}
\begin{figure}[hbt!]
	\centering
	\includegraphics[width=.49\textwidth]{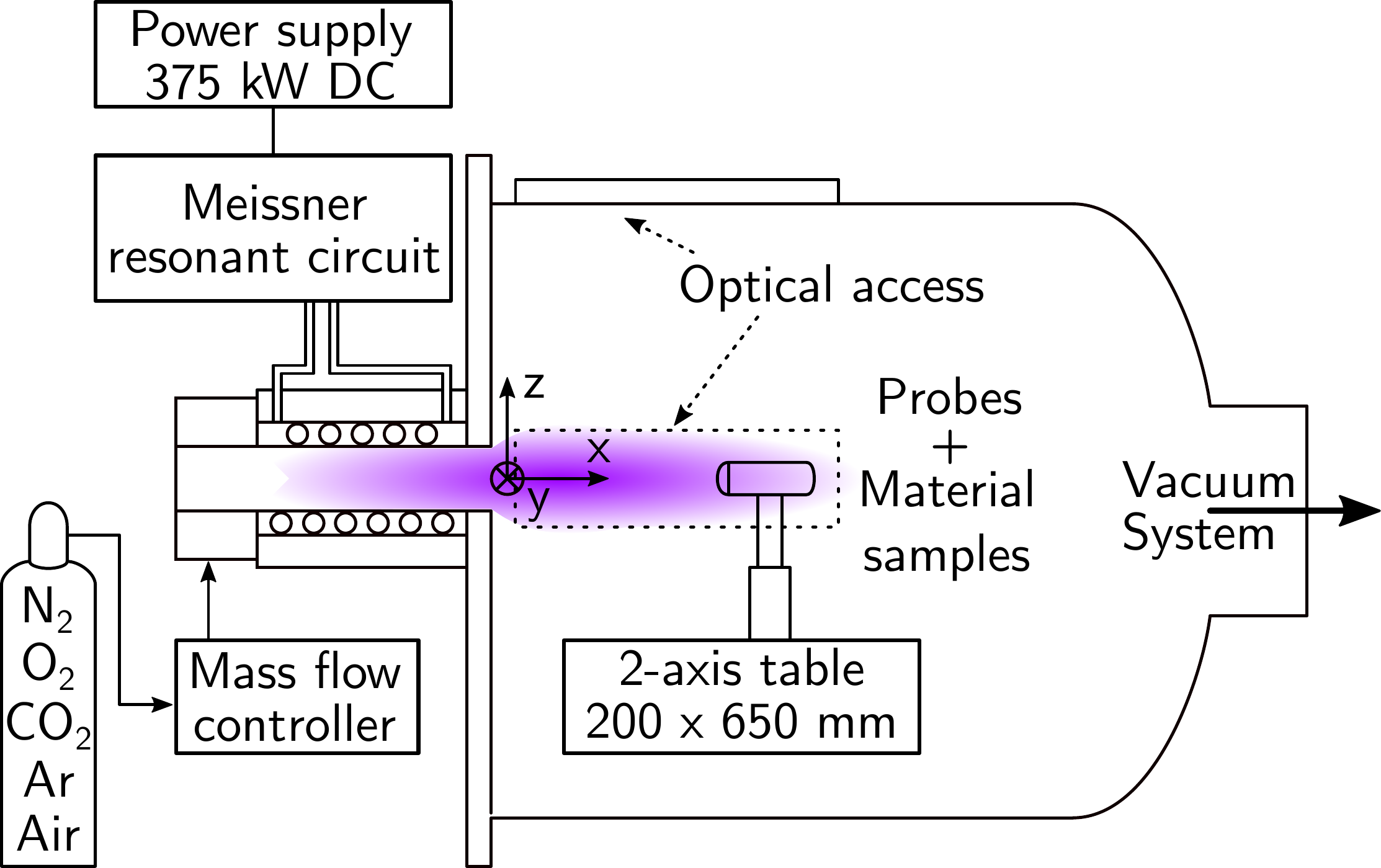}
	\caption{Schematic of plasma wind tunnel PWK3 with probe installed.}
	\label{fig:pwk3_facility_setup}
\end{figure}

The plasma wind tunnel PWK3 consists of a stainless-steel vacuum chamber with a diameter of $1.6\,\mathrm{m}$ and a length of $2\,\mathrm{m}$, as well as an inductive plasma generator (IPG3). In the case of CO\textsubscript{2} operation a convergent nozzle is attached to the plasma source, then called IPG4. A schematic of the facility is shown in Fig. \ref{fig:pwk3_facility_setup}.

The vacuum chamber is connected to a centralized vacuum system \cite{Loehle2021}. Several optical access points allow for non-intrusive diagnostics and visual monitoring. Moreover, a water-cooled probe and material sample-holder can be mounted on a numerically controlled two-axis table in the tank center, e.g., for intrusive measurements and material testing. The reference coordinate system of PWK3 is marked in Fig. \ref{fig:pwk3_facility_setup}. Its origin ($x=0\,\mathrm{mm}$) lies in the generator exit plane, but for reasons of readability it is shifted to the right in the figure.\par 
The plasma generator IPG4, connected to the tank lid, is powered by an external power supply. Together with a resonant circuit consisting of up to seven $6\,\mathrm{nF}$ capacitors and the IPG inductance coil the power supply delivers an anode power of $180\,\mathrm{kW}$ maximum. A $2.0\,\mathrm{mm}$ thick quartz tube, surrounded by a copper coil of 5.5 turns, forms the discharge channel of IPG4. The convergent nozzle attached to IPG4 has a length of $35\,\mathrm{mm}$ and a throat diameter of $50\,\mathrm{mm}$. Thus, the nozzle exit plane lies at $x=35\,\mathrm{mm}$. The coil, quartz tube and nozzle are cooled by high-pressure water.\par
\begin{figure*}[hbt!]
	\centering
	\begin{minipage}{.49\textwidth}
		\centering
		\includegraphics[width=\textwidth]{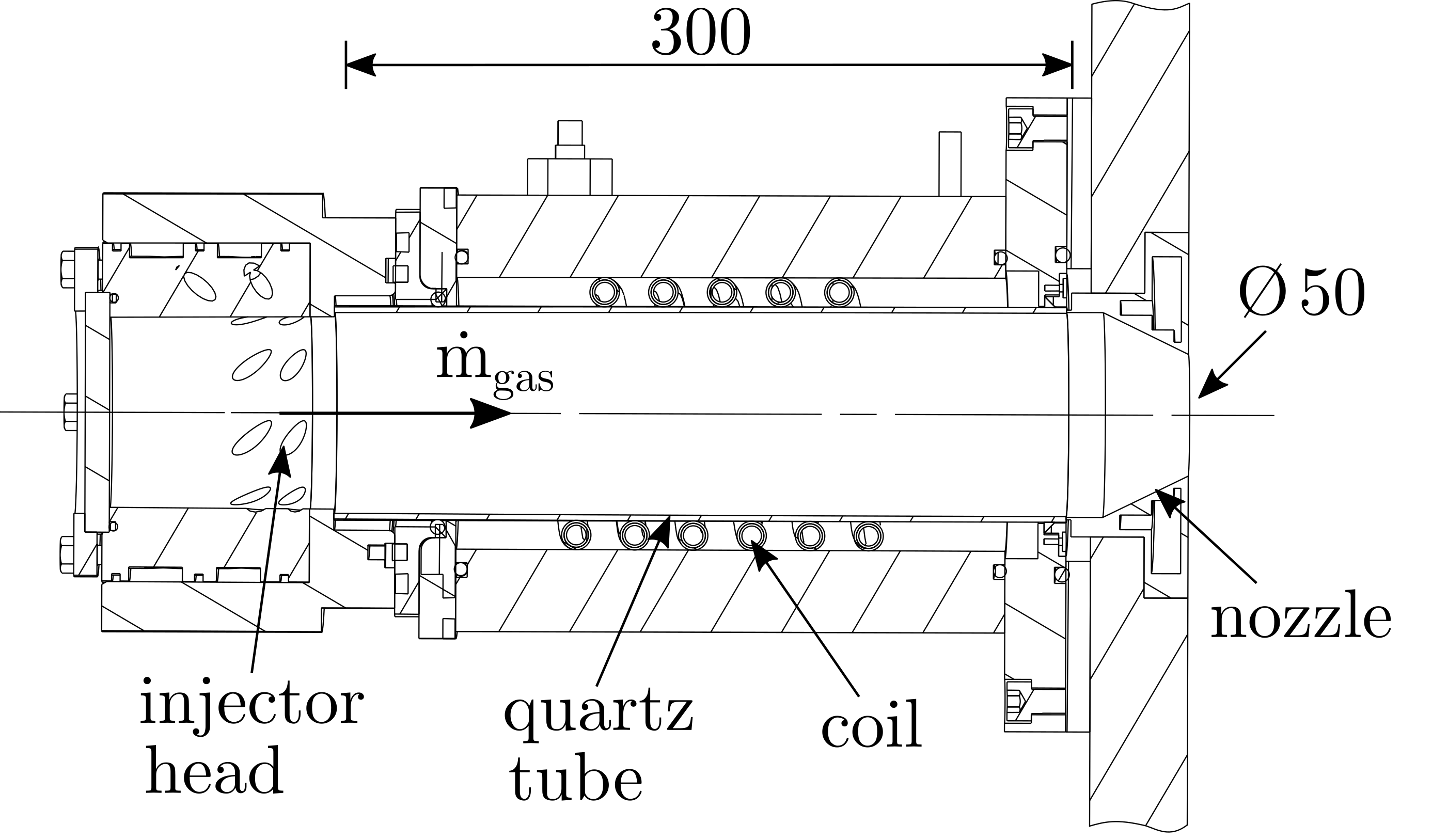}
	\end{minipage}
	\begin{minipage}{.45\textwidth}
		\centering
		\includegraphics[width=\textwidth]{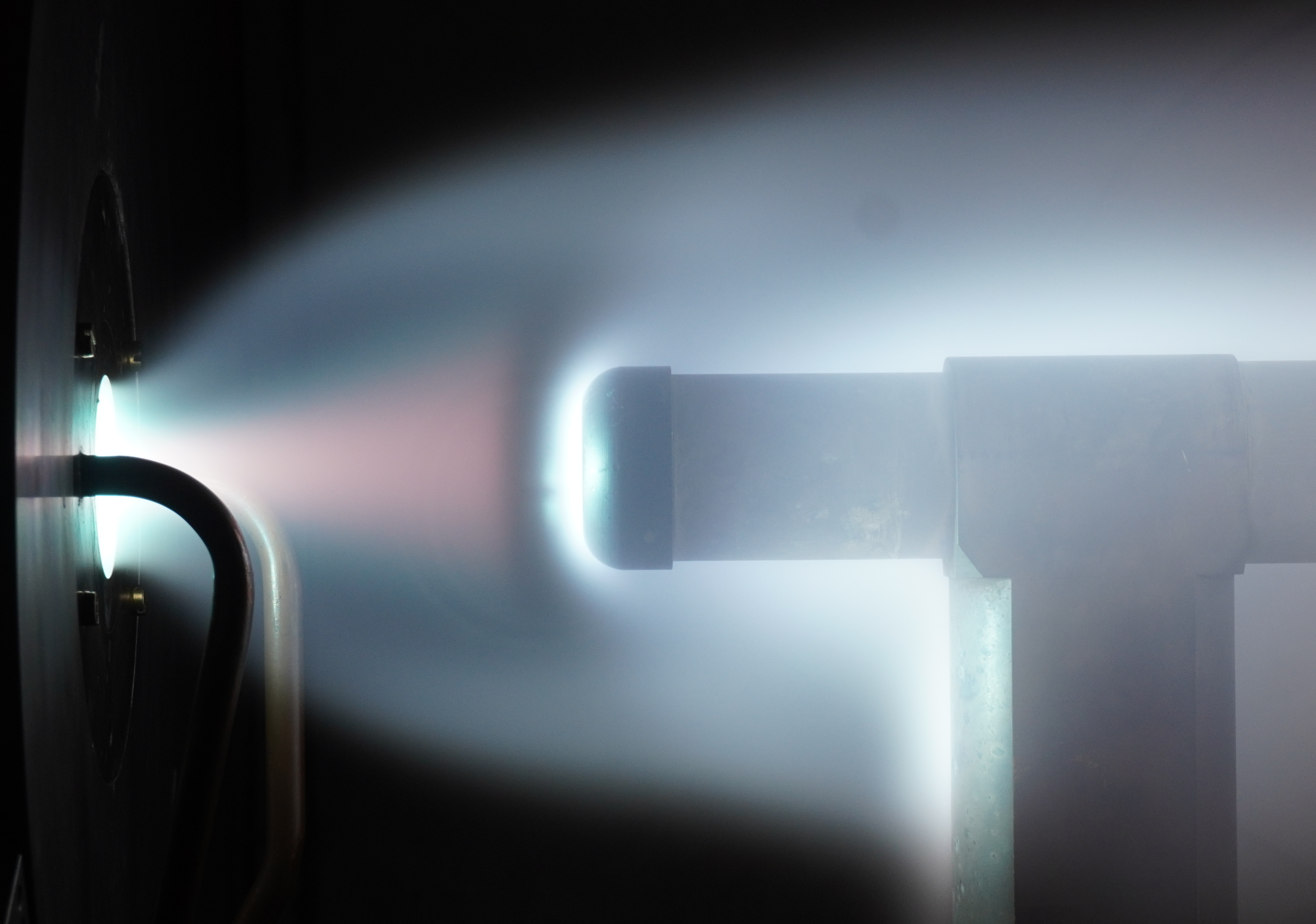}
	\end{minipage}
	\caption{Schematic of the inductive plasma generator IPG4 (left) and photograph of its CO\textsubscript{2} plasma plume during heat flux measurements at $\dot{m}_\mathrm{CO2}=2.2\,\mathrm{g/s}$, $P_\mathrm{A}=160\,\mathrm{kW}$ and $p_\mathrm{amb}=100\,\mathrm{Pa}$ in PWK3 (right).}
	\label{fig:ipg4_schematic_photo}
\end{figure*}

 
Figure \ref{fig:ipg4_schematic_photo} shows a schematic of the inductive plasma generator, as well as a photograph (Sony A6400) of a heat flux measurement in a pure CO\textsubscript{2} jet in PWK3.

The electrodeless plasma generation in IPG4 allows for the operation of PWK3 with a large variety of gases, including pure oxygen and carbon dioxide. The high plasma purity enables investigations on gas-surface interactions, e.g., for thermal protection system (TPS) materials (\cite{Pidan2005, Massuti2018, Kaiser2022}). In this work, a range of operating conditions of PWK3 with carbon dioxide is studied. A summary can be found in Tab. \ref{tab:test_parameters}.

\begin{table}[ht]
	\centering
	\caption{Range of operating conditions of PWK3/IPG4 used in this work}
	\label{tab:test_parameters}
	\begin{tabular}{llr}
		\toprule
		Parameter & Symbol & Value\\
		\cmidrule(r){1-1}\cmidrule(rl){2-2}\cmidrule(l){3-3}
		Anode power & $P_\mathrm{A}$ &$135-160\,\mathrm{kW}$\\
		Anode voltage & $U_\mathrm{A}$ &$6600-6950\,\mathrm{V}$\\
		Number of capacitors & $n_\mathrm{k}$ & 6\\
		Coil turns & $n_\mathrm{coil}$ & 5.5\\
		Operational frequency & $f$ &$520\,\mathrm{kHz}$\\
		Quartz tube thickness & $th_\mathrm{tube}$ & $2.0\,\mathrm{mm}$\\
		Attached nozzle & - & convergent\\
		Ambient pressure& $p_\mathrm{amb}$ &$30-100\,\mathrm{Pa}$\\
		Injector pressure& $p_\mathrm{inj}$ &$2855-3720\,\mathrm{Pa}$\\
		Mass flow rate [CO\textsubscript{2}]& $\dot{m}_\mathrm{CO2}$ &$2.2-4.0\,\mathrm{g/s}$\\
		\bottomrule
	\end{tabular}
\end{table}

In the context of this paper, the maximum operational range of IPG4 with regards to the mean mass-specific enthalpy is characterized. This is achieved by variation of the anode power between $135\,\mathrm{kW}$ and $160\,\mathrm{kW}$, and variation of the mass flow rate between $2.2\,\mathrm{g/s}$ and $4.0\,\mathrm{g/s}$. The limitations on that range stem from laboratory safety regulations, as well as the generator discharge stability. Except for the mass flow rate and anode power, the facility parameters are kept constant. The changes in anode voltage and injector pressures are directly connected to the change in specific energy. The tank pressure is adjusted actively by the injection of molecular nitrogen at the back of the vacuum tank. Detailed information on the investigated combinations of mass flow rate and anode power will be given in Tab. \ref{tab:test_conditions_cavity} in Section \ref{sec:results_plasma_enthalpy}.

\subsection{Plasma diagnostics}
As mentioned in Section \ref{sec:introduction}, this paper deals with spatial inhomogeneities in the mass-specific enthalpy of a carbon dioxide plasma jet, meaning the distribution of energy per particle over the radius. Consequently, the analysis is based on the comparison of the mean (bulk) enthalpy of the whole plume and locally measured values. For this purpose, two types of intrusive plasma probes, the cavity calorimeter for integral measurements and the heat flux-Pitot double probe for radially resolved measurements, are applied. Both setups and the underlying working principles of the probes are explained in the following.\par

\subsubsection{Heat flux-Pitot double probe}
\begin{figure}[hbt!]
	\centering
	\includegraphics[width=.49\textwidth]{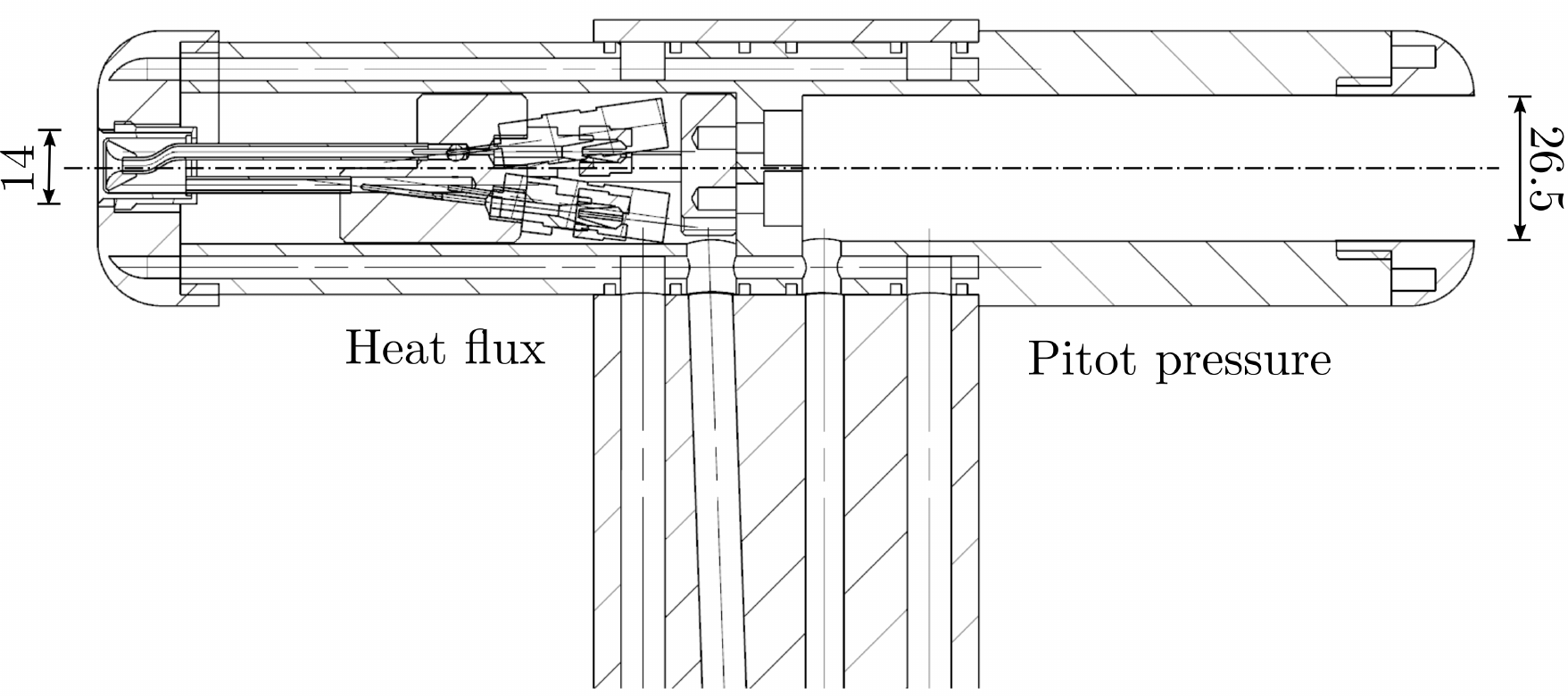}
	\caption{Sectional view of the heat flux-Pitot double probe with the $14\,\mathrm{mm}$ copper calorimeter insert on the left and the $26.5\,\mathrm{mm}$ Pitot tube on the right.}
	\label{fig:schematic_double_probe}
\end{figure}

To determine a radial profile of local mass-specific enthalpies in the CO\textsubscript{2} plasma jet of IPG4 the so-called heat flux-Pitot double probe is used. The measurements are performed at an axial distance of $x=156\,\mathrm{mm}$ from the generator exit, which equals a distance of $121\,\mathrm{mm}$ to the nozzle exit. This test position is chosen for reasons of comparison to measurements performed in the past \cite{Burghaus2023}. A schematic of the double probe is shown in Fig. \ref{fig:schematic_double_probe}.

The two-sided probe is capable of measuring the heat flux and the Pitot pressure at the test position, depending on which side of the probe is facing the plasma. Both probe heads have an outer diameter of $50\,\mathrm{mm}$. On the left side (Fig. \ref{fig:schematic_double_probe}) the heat flux on a thermally insulated copper oxide surface with a diameter of $14\,\mathrm{mm}$ is measured calorimetrically. The heat flux insert is water-cooled and the water volume flow is monitored by a Siemens Sitrans FM MAG 1100/5000 sensor system. Moreover, the temperature difference of the in- and out-flowing water $\Delta T_\mathrm{w}$ is measured by two Pt100 sensors inside the probe. Together with the known heat capacity $c_\mathrm{p,w}$, the water mass density $\rho_\mathrm{w}$ and the calorimeter surface area $A_\mathrm{cal}$, the calorimetric heat flux on the copper surface $\dot{q}_\mathrm{cal}$ can be determined as:

\begin{equation}
	\label{eq:heat_flux_double_probe}
	\dot{q}_\mathrm{cal} = \frac{\rho_\mathrm{w} \dot{V}_\mathrm{w}}{A_\mathrm{cal}} c_\mathrm{p,w} \Delta T_\mathrm{w}
\end{equation}

Opposite of the calorimeter probe head is a Pitot tube with a diameter of $26.5\,\mathrm{mm}$. The Pitot pressure is measured with a pressure gauge (MKS 122AAX-00100DBS) connected to the Pitot side of the double probe. The same gauge is used for the determination of the tank pressure at all test conditions in this work. Together, the local stagnation pressure and the calorimetrically determined heat flux allow for an approximation of the local mass-specific enthalpy $h_\mathrm{tot}$, as formulated by Marvin and Pope \cite{Marvin1967}:

\begin{equation}
	\label{eq_marvin_pope}
	h_\mathrm{tot} - h_\mathrm{w} = \frac{1}{K_\mathrm{i}}\frac{\dot{q}_\mathrm{fc}}{\sqrt{(\frac{p_\mathrm{pitot}}{R_\mathrm{eff}})}}
\end{equation}

Here, $K_\mathrm{i}$ denotes a gas specific constant, which is adopted from Zoby as $0.4337\,\mathrm{kWkg/(m^{3/2}Pa^{1/2}MJ)}$ for CO\textsubscript{2} \cite{Zoby1968, ASTM-E637}. The parameter $\dot{q}_\mathrm{fc}$ represents the fully catalytic heat flux and $R_\mathrm{eff}$ the effective nose radius, which is 2.3 times the body radius $R_\mathrm{b}=25\,\mathrm{mm}$ for the double probe used in the course of this work \cite{Marvin1967}. In this analysis, the wall enthalpy $h_\mathrm{w}$ is neglected and the measured heat flux is assumed to be fully catalytic, based on past investigations at IRS by Marynowski et al. \cite{Marynowski2014}.\par

\subsubsection{Cavity calorimeter}
\label{sec:setup_cavity_calorimeter}
\begin{figure}[hbt!]
	\centering
	\includegraphics[width=.49\textwidth]{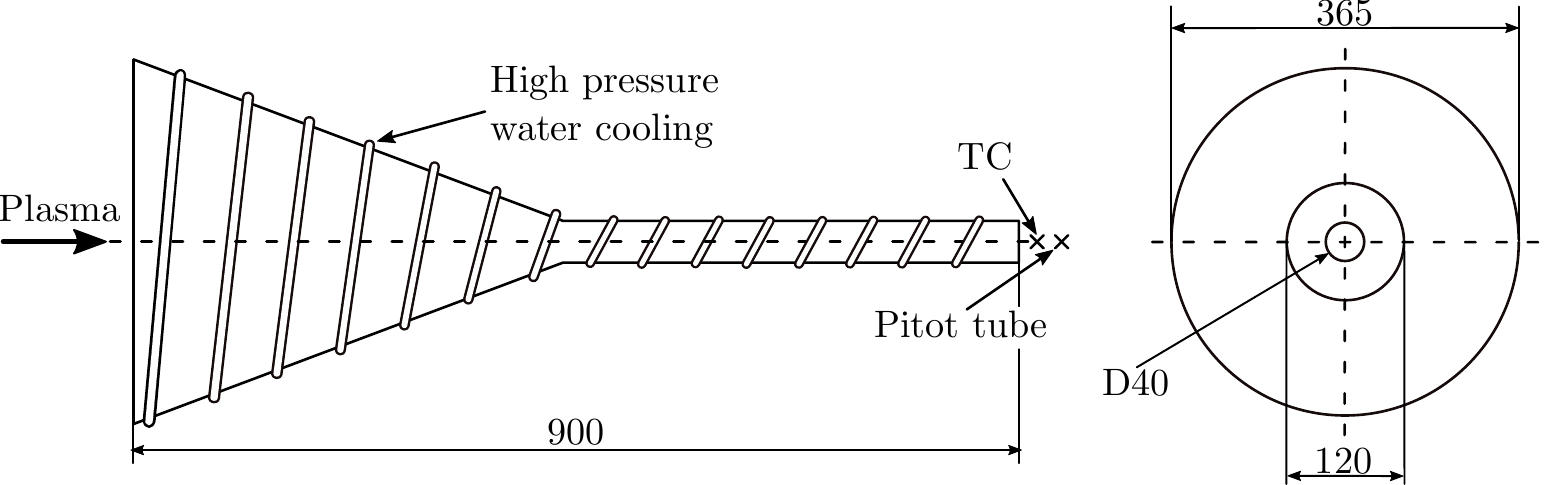}
	\caption{Schematic of the cavity calorimeter, including the placement of the thermocouple (TC) and the Pitot tube at the exit.}
	\label{fig:schematic_cavity_calorimeter}
\end{figure}

In order to measure the bulk enthalpy of the plasma jet in PWK3, a so-called cavity calorimeter was developed at the Institute of Space Systems \cite{Herdrich2002}. The basic idea behind this probe is to trap the whole plasma jet inside the calorimeter. Constant water cooling of the probe induces full relaxation of the gas temperature, chemical potential and flow velocity, allowing for the calorimetric determination of the plasma power. The cavity calorimeter is mounted to the probe holder in PWK3 and placed at a distance of $x=100\,\mathrm{mm}$ to the generator exit. This value was determined to be optimal for capturing the entire plasma jet, with no significant residual plasma flow around the probe, which would falsify the measurements, and without disturbing the discharge in the generator \cite{Herdrich2002}. Once positioned in front of the generator exit, the calorimeter was not moved anymore and the ignition was done with the probe in place already. In Fig. \ref{fig:schematic_cavity_calorimeter} a schematic of the cavity calorimeter is shown.

The cone-shaped calorimeter with a length of $900\,\mathrm{mm}$ is made out of copper. Its inlet, a circular orifice, has a diameter of $120\,\mathrm{mm}$, which is more than double the size of the IPG4 nozzle exit. The calorimeter cone is followed by a cylindrical tube of $40\,\mathrm{mm}$ diameter at the end of which the cooled gas exits the probe. All calorimeter surfaces are cooled by water in spiral copper tubes on the inside and outside. The water flow rate is monitored by a Siemens Sitrans FM MAG 1100/5000 sensor system. Moreover, the inflow and outflow temperatures of the cooling water are measured with two Pt100 sensors (not shown in Figure \ref{fig:schematic_cavity_calorimeter}). In addition to the original test setup (see \cite{Herdrich2002}) the cavity calorimeter is complemented by a Type K thermocouple (TC) and a Pitot tube, which are placed at the calorimeter end, inside the exiting gas stream. Later it will be shown that these additional diagnostics allow for the determination of the residual enthalpy content of the exiting gas and, thus, for a more accurate determination of the plasma jet power.\par

The methodology behind the data analysis is based on Herdrich \cite{Herdrich2012, Herdrich2002} and has been advanced in the course of this work. Generally, the cavity calorimeter is able to measure the total mean enthalpy of the plasma, introduced by the generator:

\begin{equation}
	\bar{h}_\mathrm{tot} = \int_{300\,\mathrm{K}}^{T} c_\mathrm{p} \mathrm{d}T + \left(h_\mathrm{chem} - \Delta H_\mathrm{f}^{0} \right) + \frac{1}{2} u_\mathrm{\infty}^2
\end{equation}

Here, $c_\mathrm{p}$ and $T$ are the specific heat capacity and the gas temperature of the plasma, respectively. The chemical potential is denoted by $h_\mathrm{chem}$. This is corrected by the standard enthalpy of formation of the gas $\Delta H_\mathrm{f}^{0}$ in order to get the enthalpy that is coupled by the plasma generator, without considering the production of CO\textsubscript{2} itself. This way, the mass-specific enthalpy is zero for a carbon dioxide gas at $300\,\mathrm{K}$, i.e., the feed gas of the generator. As the last contribution to the total enthalpy, the mass-specific kinetic energy is represented by the flow velocity $u_\mathrm{\infty}$.\par 
From the principle of operation, the cavity calorimeter measures power, not enthalpy. In this work, the total calorimetric plasma power $P_\mathrm{cal}$ is determined by combining the calorimeter, TC and Pitot tube measurements. Radiation losses of the calorimeter are assumed to be one percent, in accordance with \cite{Herdrich2002}, to stay consistent with research conducted in the past.

\begin{equation}
	P_\mathrm{cal} = 1.01\cdot P_\mathrm{cavity} + P_\mathrm{heat,exit} + P_\mathrm{kin, exit}
\end{equation}

In addition to the plasma power measured by the cavity calorimeter $P_\mathrm{cavity}$, the thermal power $P_\mathrm{heat,exit}$ and the kinetic power $P_\mathrm{kin, exit}$ left in the gas at the calorimeter exit are considered. The calorimeter power is calculated similarly to the heat flux of the double probe (Eq. \ref{eq:heat_flux_double_probe}), but for the entire plasma jet at once rather than locally:

\begin{equation}
	\label{eq:cal_mes}
	P_\mathrm{cavity} = \rho_\mathrm{W} c_\mathrm{p,W} \dot{V}_\mathrm{W}(T_\mathrm{out}-T_\mathrm{in})
\end{equation}

To estimate the thermal power at the exit, the mass flow rate $\dot{m}_\mathrm{exit}$ and the mass-specific enthalpy of the cooled gas $h_\mathrm{exit}$ must be known:

\begin{equation}
	\label{eq:heat_exit}
	P_\mathrm{heat, exit} = \dot{m}_\mathrm{exit} h_\mathrm{exit}(p_\mathrm{amb},T_\mathrm{exit})
\end{equation}

For this, thermodynamic equilibrium is assumed at the calorimeter exit. The software toolkit Cantera \cite{Cantera} is used to calculate the specific enthalpy based on the measured gas temperature at the exit. More information on these type of simulations will be given in Section \ref{sec:CO2_splitting_cantera}. Furthermore, the static pressure at the exit is set equal to the ambient pressure in the tank. Since the exit mass flow rate is not measured, it is assumed to be equal to the initial mass flow rate in IPG4. The underlying assumption here is that the plasma jet is captured completely inside the calorimeter.\par 

The kinetic power of the exiting gas completes the analysis and is determined as follows:

\begin{equation}
	\label{eq:kin_exit}
	P_\mathrm{kin, exit} = \frac{1}{2} \dot{m}_\mathrm{exit} u_\mathrm{exit}^2
\end{equation}

Here, the flow velocity of the exiting gas $u_\mathrm{exit}$ is calculated combining the measured gas temperature and the Mach number at the exit. The Mach number itself is calculated via the Rayleigh Pitot formula \cite{Anderson2001}, using the stagnation pressure measured with the Pitot tube. For subsonic conditions the simpler isentropic flow equation must be applied.\par

During the operation of plasma wind tunnel PWK3 significant power losses occur due to cooling of the quartz tube of the generator and the convergent nozzle attached to it. The IPG4 quartz tube cooling power $Q_\mathrm{tube}$ is monitored constantly for each test performed. Moreover, measurements of the nozzle cooling power $Q_\mathrm{nozzle}$ for all conditions investigated in the course of this work were performed. Both cooling losses are measured calorimetrically.
Together with the measured calorimetric plasma power, important generator/facility efficiencies can be derived. The definitions of the efficiencies are based on Dropmann and Herdrich \cite{Dropmann2013, Herdrich2012} and summarized in the following.\par

The coupling efficiency describes how much power is coupled into the working gas, compared to the power applied to the anode. This includes the calorimetric plasma power, the tube cooling and the nozzle cooling:
\begin{equation}
	\label{eq:eta_couple}
	\eta_\mathrm{couple} = \frac{P_\mathrm{cal} + Q_\mathrm{tube}+ Q_\mathrm{nozzle}}{P_\mathrm{A}}
\end{equation}

The tube cooling contains a small amount of heat generated in the IPG4 copper coil that is not coupled to the plasma and should be excluded in Eq. \ref{eq:eta_couple}. However, due to the negligible size of that coil heating (order of Watts), no correction is applied. 
The thermal efficiency states how much of the coupled power is lost due to cooling of the generator discharge channel and the convergent nozzle:
\begin{equation}
	\label{eq:eta_th}
	\eta_\mathrm{th} = \frac{P_\mathrm{cal}}{P_\mathrm{cal} + Q_\mathrm{tube}+ Q_\mathrm{nozzle}}
\end{equation}

Most importantly, the total efficiency is the ratio of the calorimetric plasma power, i.e., the power actually present in the gas in the vacuum chamber, to the anode power:
\begin{equation}
	\label{eq:eta_tot}
	\eta_\mathrm{tot} = \eta_\mathrm{couple} \cdot \eta_\mathrm{th} = \frac{P_\mathrm{cal}}{P_\mathrm{A}}
\end{equation}

Extensive studies on the PWK3 efficiencies for pure oxygen plasmas by Herdrich can be found in \cite{Herdrich2004}. The current paper represents the first measurement of the plasma power and the facility efficiencies for a carbon dioxide plasma in PWK3 (see Section \ref{sec:results_plasma_enthalpy}).

\subsection{CO\textsubscript{2} splitting in thermodynamic equilibrium}
\label{sec:CO2_splitting_cantera}
\begin{figure}[hbt!]
	\centering
	\includegraphics[width=.49\textwidth]{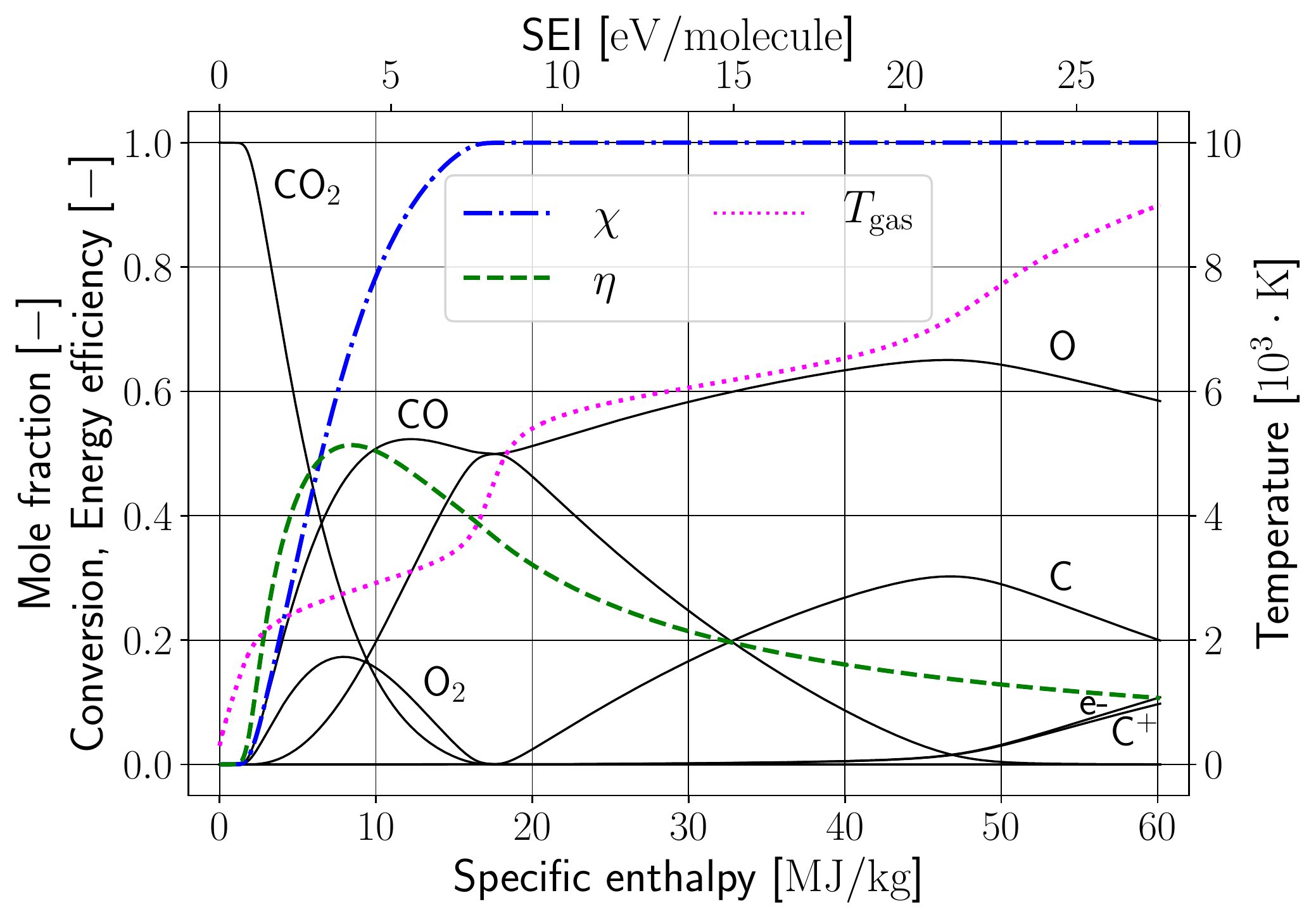}
	\caption{Composition and splitting performance of a CO\textsubscript{2} gas in thermodynamic equilibrium as a function of specific enthalpy at $2900\,\mathrm{Pa}$. The gas temperature is shown in magenta, the conversion in blue and the energy efficiency in green.}
	\label{fig:cantera_splitting}
\end{figure}

Heating carbon dioxide up to temperatures of several thousand degrees leads to decomposition of the otherwise stable molecule. This splitting process in thermochemical equilibrium is the simplest route of  CO\textsubscript{2} dissociation, but is generally limited with regards to energy efficiency as mentioned in Section \ref{sec:introduction}. In this work, a code has been  developed, based on the software toolkit Cantera \cite{Cantera}, to simulate the process of thermal carbon dioxide splitting at a given pressure. This includes the splitting performance parameters, like specific energy input (SEI), the CO\textsubscript{2} conversion $\chi$ and the energy efficiency $\eta$. The definitions of the CO\textsubscript{2} splitting performance parameters follow Snoeckx et al. \cite{Snoeckx2017}, but with adaptation to the applications in this paper. 
In the context of this work, ideal quenching without heat recovery is assumed.
The conversion of carbon dioxide is defined as the ratio of converted to the initial CO\textsubscript{2} mass:

\begin{equation}
	\chi = \frac{m_\mathrm{CO2,0} - m_\mathrm{CO2}}{m_\mathrm{CO2,0}} = \frac{m_\mathrm{CO2,converted}}{m_\mathrm{CO2,0}}
\end{equation}

with the initial and current carbon dioxide masses being $m_\mathrm{CO2,0}$ and $m_\mathrm{CO2}$, respectively. In the case of thermodynamic equilibrium CO\textsubscript{2} splitting, "current" refers to the state after heating. Consequently, the energy efficiency is defined as:

\begin{equation}
	\eta = \chi \cdot \frac{\Delta H^\mathrm{0}_\mathrm{298}}{\mathit{SEI}} = \chi \cdot \frac{\Delta H^\mathrm{0}_\mathrm{298}}{h_\mathrm{tot}} \frac{e \cdot N_\mathrm{A}}{M_\mathrm{CO2}}
\end{equation}

with the CO\textsubscript{2} conversion $\chi$ and the standard reaction enthalpy for carbon dioxide splitting $\Delta H^\mathrm{0}_\mathrm{298} = 2.93\,\mathrm{eV/molecule}$ \cite{Snoeckx2017}. In Fig. \ref{fig:cantera_splitting} the composition and splitting performance of a CO\textsubscript{2} gas in thermodynamic equilibrium are plotted. The pressure in the simulation is $2900\,\mathrm{Pa}$, representative for the injector pressure of IPG4 at the lowest mass flow rate of $2.2\,\mathrm{g/s}$ and an anode power of $160\,\mathrm{kW}$.

In the figure, the specific energy input in $\mathrm{eV/molecule}$ is plotted as a secondary x-axis for reasons of convenience. The gas composition is depicted as black lines. At approx. $16\,\mathrm{MJ/kg}$ the dissociation of carbon dioxide is completed, which is reflected in the conversion (blue) approaching $100\,\%$. Consequently, at higher enthalpies carbon monoxide decomposition and ionization, mostly of atomic carbon, takes place. The gas temperature, which is very sensitive to the static pressure, reaches approx. $3500\,\mathrm{K}$ at full CO\textsubscript{2} dissociation and rises to $9000\,\mathrm{K}$ at an enthalpy of $60\,\mathrm{MJ/kg}$. The energy efficiency reaches its maximum of approx. $51\,\mathrm{\%}$ at an enthalpy of $8.4\,\mathrm{MJ/kg}$. Again, it has to be emphasized that the CO\textsubscript{2} splitting performance is calculated under the assumption of ideal quenching without heat recovery.

\section{Results and discussion}
\label{sec:results}
As this paper strives to investigate the influence of a spatially inhomogeneous distribution of the specific energy (mass-specific enthalpy) on the CO\textsubscript{2} splitting performance, the mandatory first step is to show and quantify the spatial inhomogeneity in PWK3. In the following, this is done by using the plasma diagnostic probes introduced in Section \ref{sec:experimental_setup}. Subsequently, the influence of the measured inhomogeneity on the carbon dioxide splitting performance in IPG4 is analyzed in Section \ref{sec:influence_of_inhomogeneity}. The majority of the data analysis is performed in Python using the NumPy \cite{NumPy2020}, Matplotlib \cite{Matplotlib2007}, SciPy \cite{SciPy2020} and Pandas \cite{Pandas2010} packages.

\subsection{Plasma jet characterization}
\label{sec:results_plasma_enthalpy}

The characterization of the IPG4 plasma jet is split into two parts. First, the mean enthalpy of the entire jet, as well as the generator efficiencies, are determined with the cavity calorimeter. Five different test conditions in a wide operational range of IPG4 are analyzed. Second, for one of these conditions (CO2\#01b) the radial distribution of local mass-specific enthalpy values is measured with the heat flux-Pitot double probe.

\subsubsection{Bulk enthalpy \& generator efficiencies}
In the past, extensive studies on the bulk enthalpy of oxygen plasmas in PWK3 were performed by Herdrich using the cavity calorimeter \cite{Herdrich2002, Herdrich2004}. The experiments conducted in the frame of this paper represent the first attempt at measuring the mean enthalpy for carbon dioxide. Moreover, the determination of the nozzle cooling power for IPG4 is a novelty. The operating parameters of the corresponding test conditions are summarized in Tab. \ref{tab:test_conditions_cavity}. Parameters that are not part of the table are the same as in Tab. \ref{tab:test_parameters} for all experiments. As stated before, the anode voltage and the injector pressure are not target parameters, but their changes are the result of altering the mass flow rate and the anode power.

\begin{table}[ht]
	\centering
	\caption{Summary of test conditions for the CO\textsubscript{2} cavity calorimeter experiments in PWK3 at varying specific anode powers $\mathit{\bar{h}_\mathrm{A}}$.}
	\label{tab:test_conditions_cavity}
	\begin{threeparttable}
		\begin{tabular}{lrrrr}
			\toprule
			Condition & $P_\mathrm{A}\,\mathrm{[kW]}$ & $\dot{m}_\mathrm{CO2}\,\mathrm{[g/s]}$ & $\mathit{\bar{h}_\mathrm{A}}\,\mathrm{[MJ/kg]}$ & $p_\mathrm{amb}\,\mathrm{[Pa]}$\tnote{a}\\
			\cmidrule(r){1-1}\cmidrule(rl){2-2}\cmidrule(rl){3-3}\cmidrule(rl){4-4}\cmidrule(l){5-5}
			CO2\#01a & $160$ & 2.2 & 72.73 & 27 \\
			CO2\#01b & $160$ & 2.2 & 72.73 & 83\\
			CO2\#02 & $160$ & 3.0 & 53.33 & 92 \\
			CO2\#03 & $150$ & 3.5 & 42.86 & 97 \\
			CO2\#04 & $135$ & 4.0 & 33.75 & 102\\
			\bottomrule
		\end{tabular}
		\begin{tablenotes}
			\item[a]{Due to placement of the cavity calorimeter in front of the generator exit, the ambient pressure is not representative of the free-stream conditions.}
		\end{tablenotes}
	\end{threeparttable}
\end{table}

The five test conditions are labeled as CO2\#01a/b-CO2\#04, differing in anode power, mass flow rate and/or tank pressure. The higher the number, the lower the ratio of applied anode power $P_\mathrm{A}$ and mass flow rate $\dot{m}_\mathrm{CO2}$, defined as $\mathit{\bar{h}_\mathrm{A}}$:

\begin{equation}
	\bar{h}_\mathrm{A} = \frac{P_\mathrm{A}}{\dot{m}_\mathrm{CO2}}
\end{equation}

The specific anode power $\mathit{\bar{h}_\mathrm{A}}$ is by definition a representation of the facility operating conditions. Consequently, it does not account for losses and is to be distinguished from the mean specific energy coupled into the plasma:
\begin{equation}
	\bar{h}_\mathrm{tot} = \eta_\mathrm{tot} \cdot \bar{h}_\mathrm{A}
\end{equation}

\begin{figure}[hbt!]
	\centering
	\includegraphics[width=.49\textwidth]{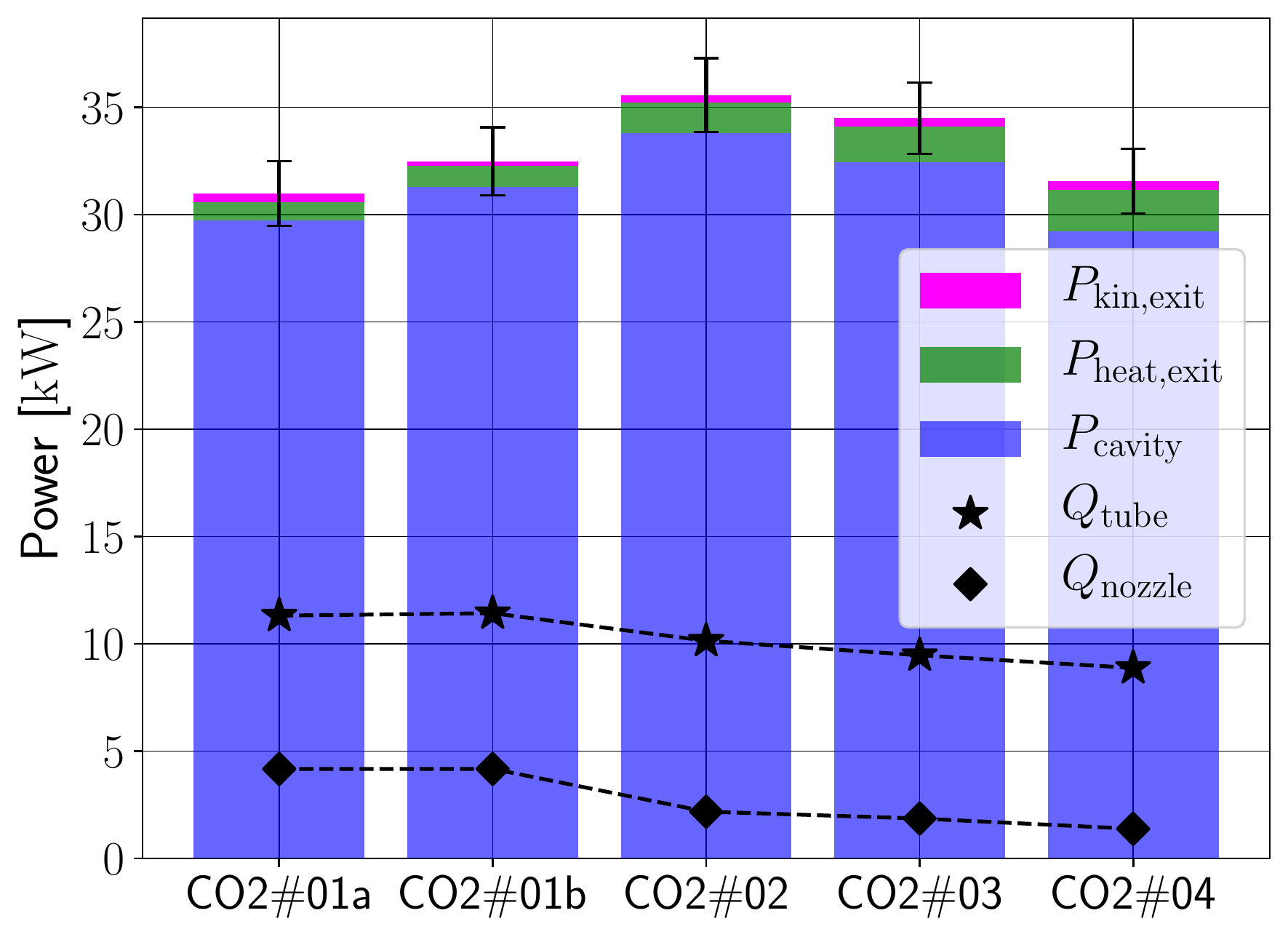}
	\caption{Measured calorimetric plasma powers for the investigated conditions. Partitions of the thermal and kinetic exit powers are shown in green and magenta, respectively. The corresponding tube and nozzle cooling powers are indicated with black stars and diamonds, respectively.}
	\label{fig:cavity_calorimeter_powers}
\end{figure}

The first two conditions are the same with regards to the generator operating parameters, but differ in the tank pressure, and are distinguished by an additional lowercase letter. The tank pressure at CO2\#01a is at minimum ($30\,\mathrm{Pa}$), while for the other conditions molecular nitrogen is injected at the back of the vacuum tank (cp. Section \ref{sec:experimental_facility}). The target pressure for CO2\#01b - CO2\#04 was $100\mathrm{Pa}$, but due to a limited test time an adjustment of the nitrogen mass flow for each condition was not possible. Moreover, capturing the plasma jet inside the cavity calorimeter influences the ambient pressure measured at the tank wall and it is not fully representative of the free-stream conditions. Nevertheless, since the ambient pressures were high enough to constrict the plasma jet to diameters smaller than the calorimeter opening, the slightly lower tank pressure is not believed to have significant influence on the cavity calorimeter measurements. In Fig. \ref{fig:cavity_calorimeter_powers}, the measured calorimetric plasma powers as well as the tube and nozzle cooling losses for all five conditions are illustrated. 

The power directly measured with the cavity calorimeter (Eq. \ref{eq:cal_mes}]) is shown in blue, while the thermal (Eq. \ref{eq:heat_exit}) and the kinetic (Eq. \ref{eq:kin_exit}) powers of the gas exiting the probe are shown in green and magenta, respectively. The tube and nozzle cooling powers are plotted in black. With regards to the nozzle cooling, the first two test conditions are treated as one, since no difference could be observed. The measured calorimetric plasma powers are in the range of approx. $30-35\,\mathrm{kW}\,$ for all conditions. It is evident that the cavity calorimeter is able to capture most of the inherent plasma power, and only little power is left in the gas exiting the device. Nevertheless, neglecting the remaining enthalpy in the outflowing gas causes an error of up to $7\,\%$ in the case of high mass flow rates, which justifies the improvements made to the test setup. Comparing CO2\#01a and CO2\#01b shows the influence of raising the ambient pressure. Not only is the measured power higher at elevated pressure, proving that capturing the whole plasma jet is not possible at minimum pressure, but also the partitions change. While the kinetic exit power decreases with increasing ambient pressure, the temperature at the exit increases. This is most likely due to the lower expansion at elevated pressure and, thus, less conversion of heat into kinetic energy. The condition CO2\#02 shows the highest calorimetric plasma power. With further decreasing anode power, the calorimetric plasma power decreases, as expected. The cooling losses are significantly lower for higher mass flow rates. Reason for this is the lower energy per particle, leading to lower temperatures in the discharge. Moreover, during the experiments it could be observed that the plasma jet decouples from the nozzle with higher mass flow rates. The exact correlation between injector pressure and jet expansion is not investigated in this paper, but can be seen as a factor for the nozzle cooling reduction.\par 

\begin{figure}[hbt!]
	\centering
	\includegraphics[width=.49\textwidth]{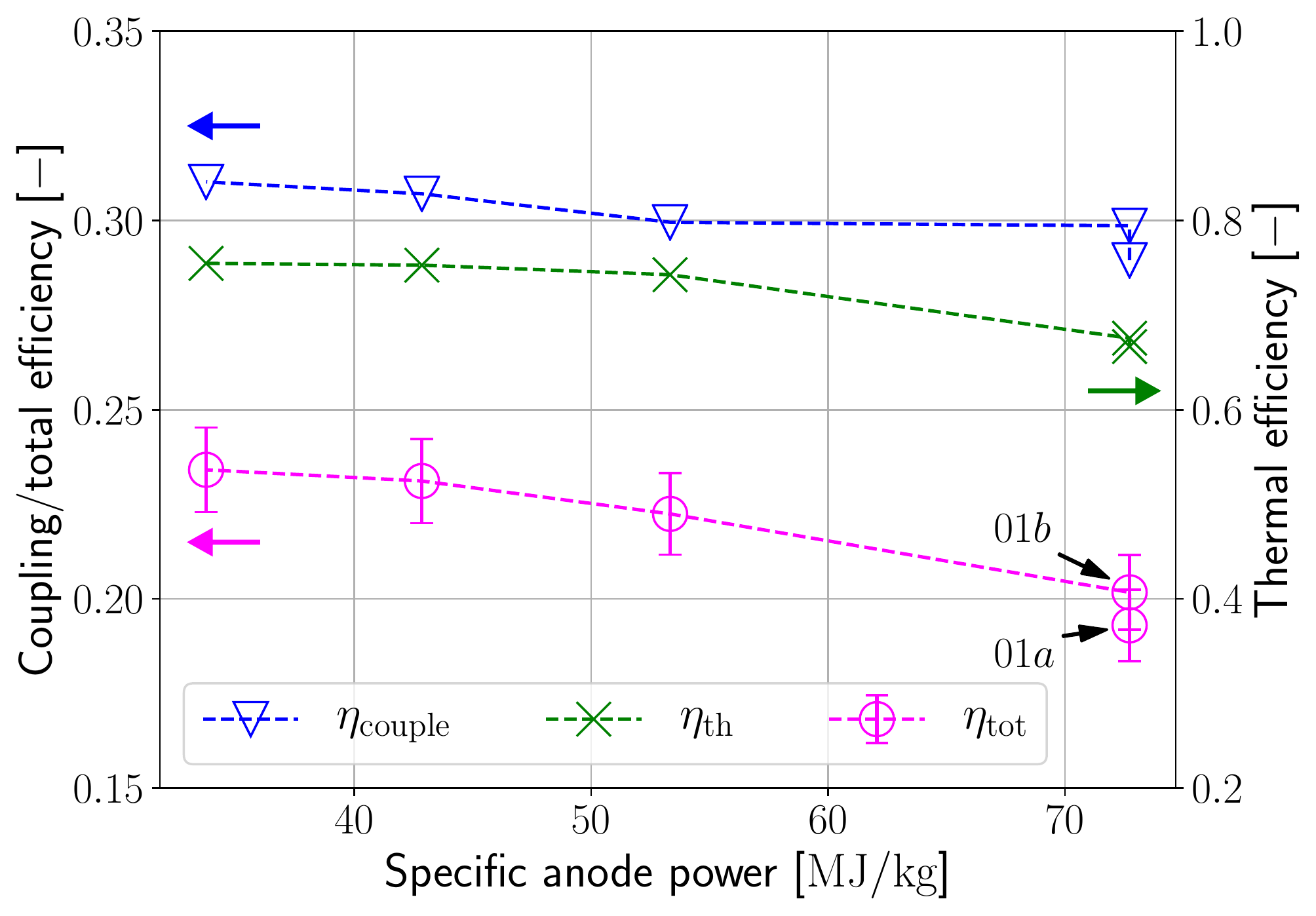}
	\caption{Measured coupling, thermal and total efficiencies plotted over the specific anode power $\mathit{\bar{h}_\mathrm{A}}$, i.e., the ratio of anode power and mass flow rate, for test conditions CO2\#01 (right) to CO\#04 (left).}
	\label{fig:cavity_calorimeter_efficiencies}
\end{figure}

Following the methodology in Section \ref{sec:setup_cavity_calorimeter}, Eqs. \ref{eq:eta_couple}-\ref{eq:eta_tot} are applied to determine relevant efficiencies of the plasma generator IPG4 and the PWK3 facility as a whole. The coupling, thermal and total efficiencies of all five test conditions are plotted over the specific anode power in Fig. \ref{fig:cavity_calorimeter_efficiencies}.

The thermal efficiency lies between 60 and 80 percent  for all conditions. The coupling efficiency is rather low ($<32\,\%$) and decreases with higher specific energies. Consequently, the total efficiency of IPG4 for pure CO\textsubscript{2} plasma flows ranges from approx. $20$ to $25$ percent, also decreasing with specific energy. The two adjacent points at the highest $\mathit{\bar{h}_\mathrm{A}}$, i.e., CO2\#01a and CO2\#01b, show a jump in efficiency when increasing the ambient pressure from $30\,\mathrm{Pa}$ to $83\,\mathrm{Pa}$. This justifies the use of elevated pressure in all experiments in order to capture the whole plasma jet inside the cavity calorimeter and avoid residual flow around the device. In contrast, the thermal efficiency is not affected by the tank pressure, indicating that the discharge in the generator itself is unaltered. The measurement results of the cavity calorimeter experiments are summarized in Tab. \ref{tab:results_cavity}. Here, the bulk enthalpy $\bar{h}_\mathrm{tot}$ is introduced as the ratio of calorimetrically measured plasma power and the total mass flow rate, assuming that the entire mass flux passes through the cavity calorimeter.

\begin{table*}[ht]
	\centering
	\caption{Summary of  cavity calorimeter results for CO\textsubscript{2} at varying specific energy input.}
	\label{tab:results_cavity}
	\begin{tabular}{lrrrrr}
		\toprule
		Condition & $Q_\mathrm{tube}\,\mathrm{[kW]}$ & $Q_\mathrm{nozzle}\,\mathrm{[kW]}$ & $P_\mathrm{cal}\,\mathrm{[kW]}$ & $\bar{h}_\mathrm{tot}\,\mathrm{[MJ/kg]}$ & $\eta_\mathrm{tot}\,\mathrm{[-]}$\\
		\cmidrule(r){1-1}\cmidrule(rl){2-2}\cmidrule(rl){3-3}\cmidrule(rl){4-4}\cmidrule(rl){5-5}\cmidrule(l){6-6}
		CO2\#01a & 11.31 & 4.17 & $30.98\pm 1.52$ & $14.08\pm 0.70$ & 0.193 \\
		CO2\#01b & 11.43 & 4.17 & $32.48\pm 1.59$ & $14.76\pm 0.74$ & 0.202 \\
		CO2\#02 &  10.14 & 2.18 & $35.57\pm 1.72$ & $11.86\pm 0.58$ & 0.222 \\
		CO2\#03 &  9.46 & 1.86 & $34.49\pm 1.66$ & $9.85\pm 0.48$ & 0.231 \\
		CO2\#04 &  8.88 & 1.38 & $31.56\pm 1.51$ & $7.89\pm 0.38$ & 0.234 \\
		\bottomrule
	\end{tabular}
\end{table*}

The measured bulk enthalpies cover an interesting range regarding thermal CO\textsubscript{2} splitting (Fig. \ref{fig:cantera_splitting}). While the high power condition CO2\#01b is expected to just about reach full conversion ($97\,\%$) at a medium high energy efficiency of $42\,\%$, condition CO2\#04 with a bulk enthalpy of $7.89\,\mathrm{MJ/kg}$ is near the optimum efficiency of $51\,\%$ at lower conversion of $63\,\%$. \par

For oxygen plasma in PWK3, Herdrich determined a total efficiency of $22\,\%$ at $\mathit{\bar{h}_\mathrm{A}}=47.7\,\mathrm{MJ/kg}$ using a 5.5 turn IPG coil and four capacitors in the resonant circuit. \cite{Herdrich2002}. This corresponds well to the measured efficiencies for carbon dioxide in the course of this work. Moreover, Herdrich observed the same trend of higher total efficiencies with lower specific anode powers, reaching up to $35\,\%$ at $\mathit{\bar{h}_\mathrm{A}}=19.3\,\mathrm{MJ/kg}$ \cite{Herdrich2002}.

\subsubsection{Radial specific enthalpy distribution}

\begin{figure}[hbt!]
	\centering
	\includegraphics[width=.49\textwidth]{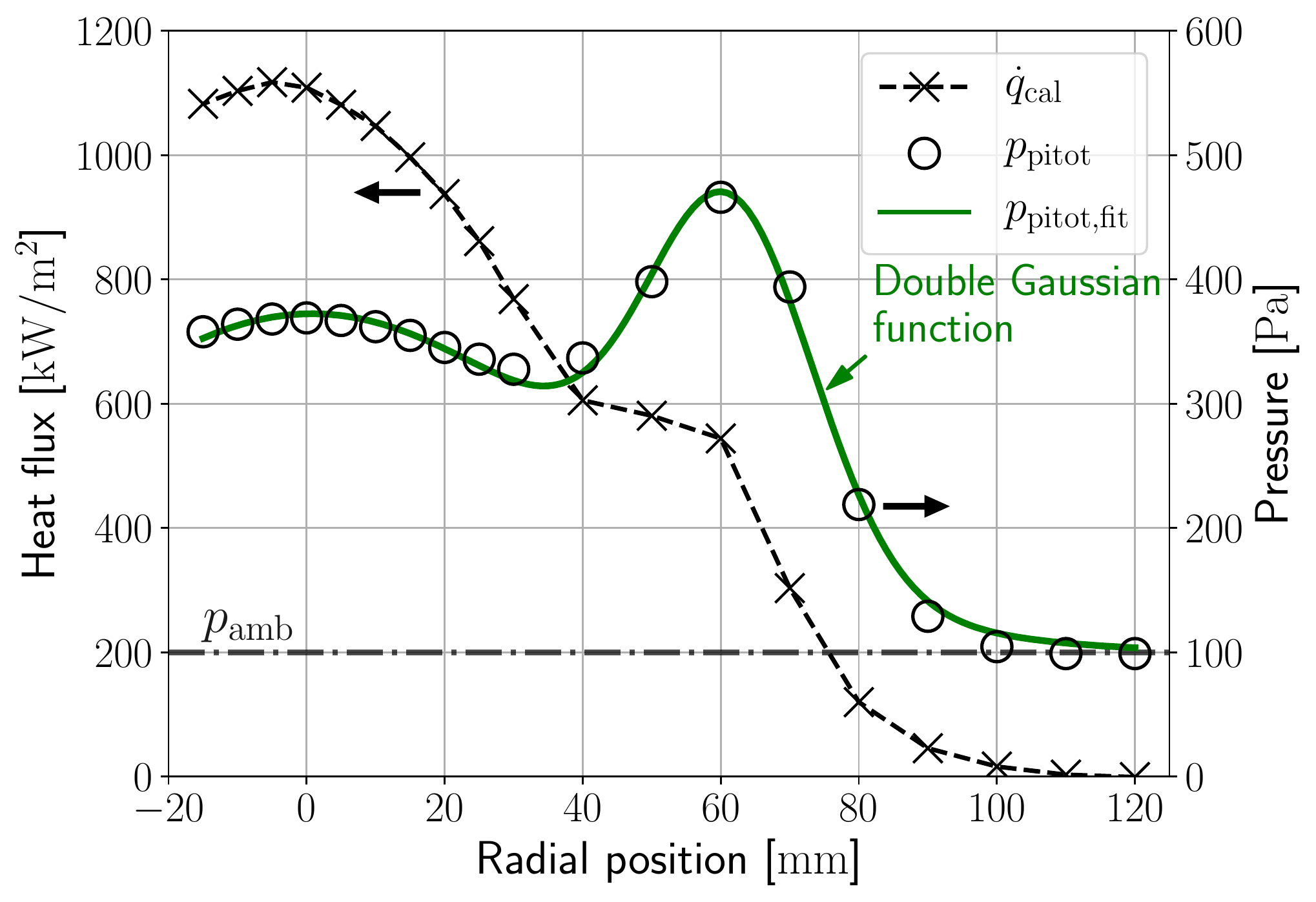}
	\caption{Radial profiles of local heat flux and Pitot pressure for the condition CO2\#01b at $x=156\,\mathrm{mm}$. A double Gaussian distribution (green line) is fitted to the Pitot values. The tank pressure of $100\,\mathrm{Pa}$ is indicated by a horizontal line.}
	\label{fig:double_probe_hf_pitot}
\end{figure}

In order to quantify the spatial specific energy inhomogeneity in the IPG4 CO\textsubscript{2} plasma jet, local measurements of the mass-specific enthalpy were conducted at an axial position of $x=156\,\mathrm{mm}$ for one condition, i.e., CO2\#01b, with the heat flux-Pitot double probe. The tank pressure was set to $100\,\mathrm{
Pa}$ for all tests. In Fig. \ref{fig:double_probe_hf_pitot} the measured heat flux, referred to the $50\,\mathrm{mm}$ probe geometry, as well as the Pitot pressure over the radius are shown. Due to their small size, the error bars are not drawn for reasons of readability. On the centerline, the relative errors are $4\,\%$ for the heat flux and $2\,\%$ for the Pitot pressure.

Slightly left of the plasma jet centerline, a maximum heat flux of $ 1117\,\mathrm{kW/m^2}$ is measured by the calorimetric insert. The small offset of the peak position to $r=0\,\mathrm{mm}$ is due to unavoidable inaccuracies in manual probe placement. Towards the plasma edge, the heat flux decreases steadily, with a notable plateau between $r=40\,\mathrm{mm}$ and $r=60\,\mathrm{mm}$. The Pitot measurement shows a prominent off-centered global peak of $466\,\mathrm{Pa}$ at $60\,\mathrm{mm}$ and a smaller local peak on the plasma jet centerline. It is believed that the off-centered global peak originates from the tangential gas injection in IPG4, which stabilizes the plasma, leading to an increased static pressure at the discharge chamber wall \cite{Herdrich2002}. The radial position of the peak is strongly influenced by the jet expansion and depends on the axial measurement position and the ambient pressure in the tank. The Pitot profile can be described by a double Gaussian distribution, which is drawn as a green line and follows the equation:
\begin{equation}
	\label{eq:double_gauss}
	f(r) = a_1 \exp \left( -\frac{(r-b_1)^2}{2c_1^2}\right) + a_2 \exp \left( -\frac{(r-b_2)^2}{2c_2^2}\right) + d
\end{equation}
with $a_1=2.72\cdot10^2\,\mathrm{Pa}$ , $b_1=7.9\cdot10^{-4}\,\mathrm{m}$, $c_1=4.06\cdot10^{-2}\,\mathrm{m}$, $a_2=2.79\cdot10^{2}\,\mathrm{Pa}$, $b_2=6.17\cdot10^{-2}\,\mathrm{m}$, $c_2=1.2\cdot10^{-2}\,\mathrm{m}$ and $d=100\,\mathrm{Pa}$. This characteristic shape will be used in Section \ref{sec:influence_of_inhomogeneity} for the determination of the local mass flux profile.

\begin{figure}[hbt!]
	\centering
	\includegraphics[width=.49\textwidth]{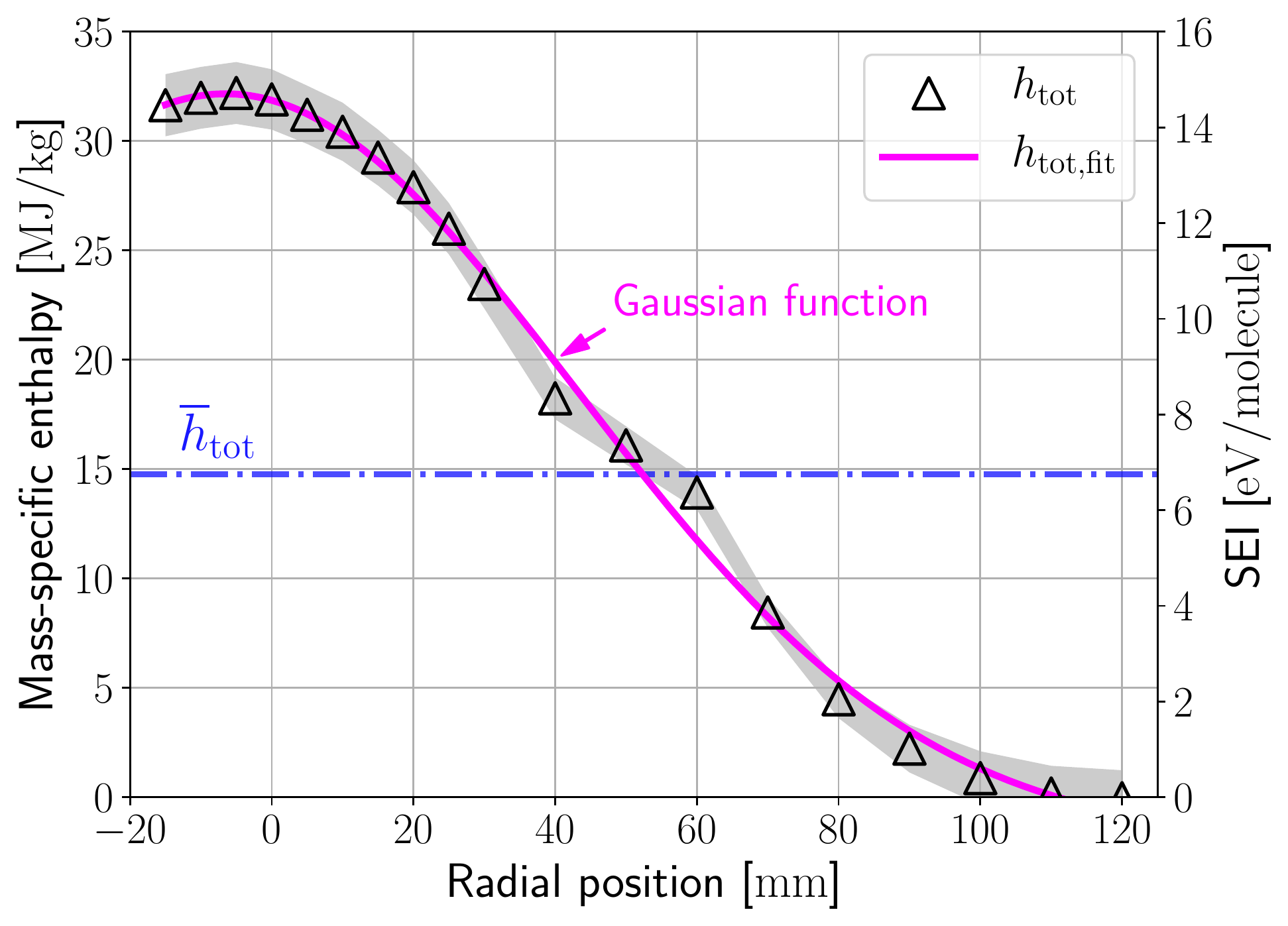}
	\caption{Radial distribution of the mass-specific enthalpy (black triangles) for the condition CO2\#01b at $x=156\,\mathrm{mm}$, with uncertainty region in gray. The mean enthalpy is indicated as a horizontal blue line. A Gaussian function (magenta) is fitted to the enthalpy data.}
	\label{fig:double_probe_enthalpy}
\end{figure}

Based on the measurements of heat flux and Pitot pressure at condition CO2\#01b, the radial distribution of the mass-specific enthalpy can be calculated, using Eq. \ref{eq_marvin_pope}. Figure \ref{fig:double_probe_enthalpy} presents the locally measured values. Here, the uncertainty region is indicated by a gray shadow.

The mass-specific enthalpy shows a bell-shaped distribution with a maximum value of $32.2\,\mathrm{MJ/kg}$ near the centerline and negligible enthalpies at the plasma edge. Based on the simulation of thermal splitting (Fig. \ref{fig:cantera_splitting}) full dissociation of CO\textsubscript{2} is expected for the inner plasma jet region, i.e., radial positions of $r<\,\mathrm{60mm}$. The highest energy efficiencies of approx. $ 50\,\%$ are estimated to be around $r=\,\mathrm{70mm}$. The mean mass-specific enthalpy of $\bar{h}_\mathrm{tot}=14.76\,\mathrm{MJ/kg}$, measured with the cavity calorimeter for CO2\#01b, is depicted as a horizontal blue line. 
Moreover, a Gaussian function, drawn in magenta, is fitted to the enthalpy data. The fit function is in good agreement with the measurements and lies withing the uncertainty region for nearly all points. The fit function follows the equation:
\begin{equation}
	f(r) = a \exp \left( -\frac{(r-b)^2}{2c^2}\right) + d
\end{equation}
with $a=3.42\cdot10^{7}\,\mathrm{J/kg}$, $b=-6.53\cdot10^{-3}\,\mathrm{m}$, $c=4.93\cdot10^{-2}\,\mathrm{m}$ and $d=-2.01\cdot10^{6}\,\mathrm{J/kg}$. \par
The probe measurements show a strong spatial inhomogeneity in specific enthalpy at the test position. On the plasma jet centerline, as well as in the outer parts of the plume, the enthalpy deviates significantly from the mean value. In the following, the influence of this spatial inhomogeneity in specific energy input on the total CO\textsubscript{2} splitting performance will be investigated.

\subsection{The influence of inhomogeneity on CO\textsubscript{2} splitting performance}
\label{sec:influence_of_inhomogeneity}

In this section, the influence of spatial enthalpy inhomogeneity on the carbon dioxide conversion and, thus, the energy efficiency of CO\textsubscript{2} splitting is examined by the example of the test condition CO2\#01b. To do so, two cases are constructed: First, a theoretical scenario where the mass-specific enthalpy is constant over radius, i.e., at all positions in the plasma jet the enthalpy equals the mean value of $\bar{h}_\mathrm{tot}=14.76\,\mathrm{MJ/kg}$. Second, the real scenario, where mass flux and enthalpy are inhomogeneously distributed over the plasma jet radius, leading to the mass-specific enthalpy profile in Fig. \ref{fig:double_probe_enthalpy}. Although the probe measurements are conducted in the vacuum tank after plasma expansion through the convergent nozzle, it is assumed that the measured specific energy distribution is representative of the inhomogeneities during plasma generation, and thus CO\textsubscript{2} conversion, in the discharge region of the plasma source. This is supported by numerical simulations by Vasil'evskii et al. of an air flow in an inductively coupled plasma (ICP) source similar to the one used in the course of this work \cite{Vasilevskii2023}. The simulations show that the inhomogeneities in mass-specific enthalpy are introduced during plasma generation in the discharge region already, before being expanded in the vacuum tank.  Thus, the measurements with the double probe in PWK3 are a measure of the inhomogeneous plasma generation process in the generator itself.\par 

The CO\textsubscript{2} conversion is in both cases, homogeneous and inhomogeneous, calculated using Cantera under the assumption of thermodynamic equilibrium (cp. Section \ref{sec:CO2_splitting_cantera}). For IPG4, this is a good approximation, at least for the plasma generation phase inside the generator, as shown in recent work \cite{Burghaus2023}. In the homogeneous case, the carbon dioxide splitting performance can be directly read from Fig. \ref{fig:cantera_splitting}. In the case of inhomogeneous enthalpy distribution, the mass distribution in the plasma jet must be known in order to calculate the integral splitting performance. Hence, the mass flux distribution at the test position in the CO\textsubscript{2} plasma jet is determined in the following.

\subsubsection{Mass flux determination}
\label{sec:mass_flex_determination}
To calculate the integral splitting performance in the inhomogeneous case, the mass flux (mass flow rate density) is required, since the measured enthalpy is mass-specific and does not reveal the enthalpy distribution per mass. The mass flux, in cylindrical coordinates ($r, \varphi$), is defined as
\begin{equation}
	\label{eq:mass_flux_definition}
	j_\mathrm{m} (r, \varphi) = \frac{\mathrm{d} \dot{m}}{\mathrm{d} A}
\end{equation}
with the mass flow rate $\dot{m}$ and the plasma jet cross-sectional area $A$. Unfortunately, there is no way to directly measure the local mass flux, but the distribution can be reconstructed using the plasma probe measurements presented earlier. In particular, three primary criteria must be met by a suitable mass flux profile:
\begin{enumerate}
	\item At the plasma edge ($R_\mathrm{max}$) the mass flux is zero:
	\begin{equation}
		\label{eq:mflux_boundary}
		j_\mathrm{m} (R_\mathrm{max}, \varphi) = 0
	\end{equation}
	\item The mass flux integrated over the plasma jet cross section equals the total mass flow rate:
	\begin{equation}
		\label{eq:mass_integral}
		\dot{m} = \int_{0}^{2 \pi} \int_{0}^{R_\mathrm{max}} j_\mathrm{m}(r, \varphi) r \,\mathrm{d} r \,\mathrm{d} \varphi
	\end{equation}
	\item The product of mass flux and specific enthalpy integrated over the plasma jet cross-sectional area is equal to the total calorimetric plasma power:
	\begin{equation}
		\label{eq:power_integral}
		P_\mathrm{cal} = \int_{0}^{2 \pi} \int_{0}^{R_\mathrm{max}} h_\mathrm{tot}(r, \varphi) j_\mathrm{m}(r, \varphi) r \,\mathrm{d} r \,\mathrm{d} \varphi
	\end{equation}
\end{enumerate}

For CO2\#01b, the total mass flow rate is known to be $\dot{m}=~2.2\,\mathrm{g/s}$. Moreover, the calorimetric plasma power $P_\mathrm{cal}=~32.48\,\mathrm{kW}$ was measured with the cavity calorimeter (Tab. \ref{tab:results_cavity}). In addition, the radial distribution of the mass specific enthalpy $ h_\mathrm{tot}(r, \varphi)$ was determined with the double probe (Fig. \ref{fig:double_probe_enthalpy}). In the course of this analysis, the Gaussian fit function of the measured enthalpies $h_\mathrm{tot,fit}$ is used instead of the single data points, serving as a continuous function in the calculations. Moreover, rotational symmetry is assumed inside the IPG4 plasma jet:
\begin{equation}
	h_\mathrm{tot} \neq f(\varphi), \,j_\mathrm{m} \neq f(\varphi)
\end{equation}

\begin{figure}[hbt!]
	\centering
	\includegraphics[width=.49\textwidth]{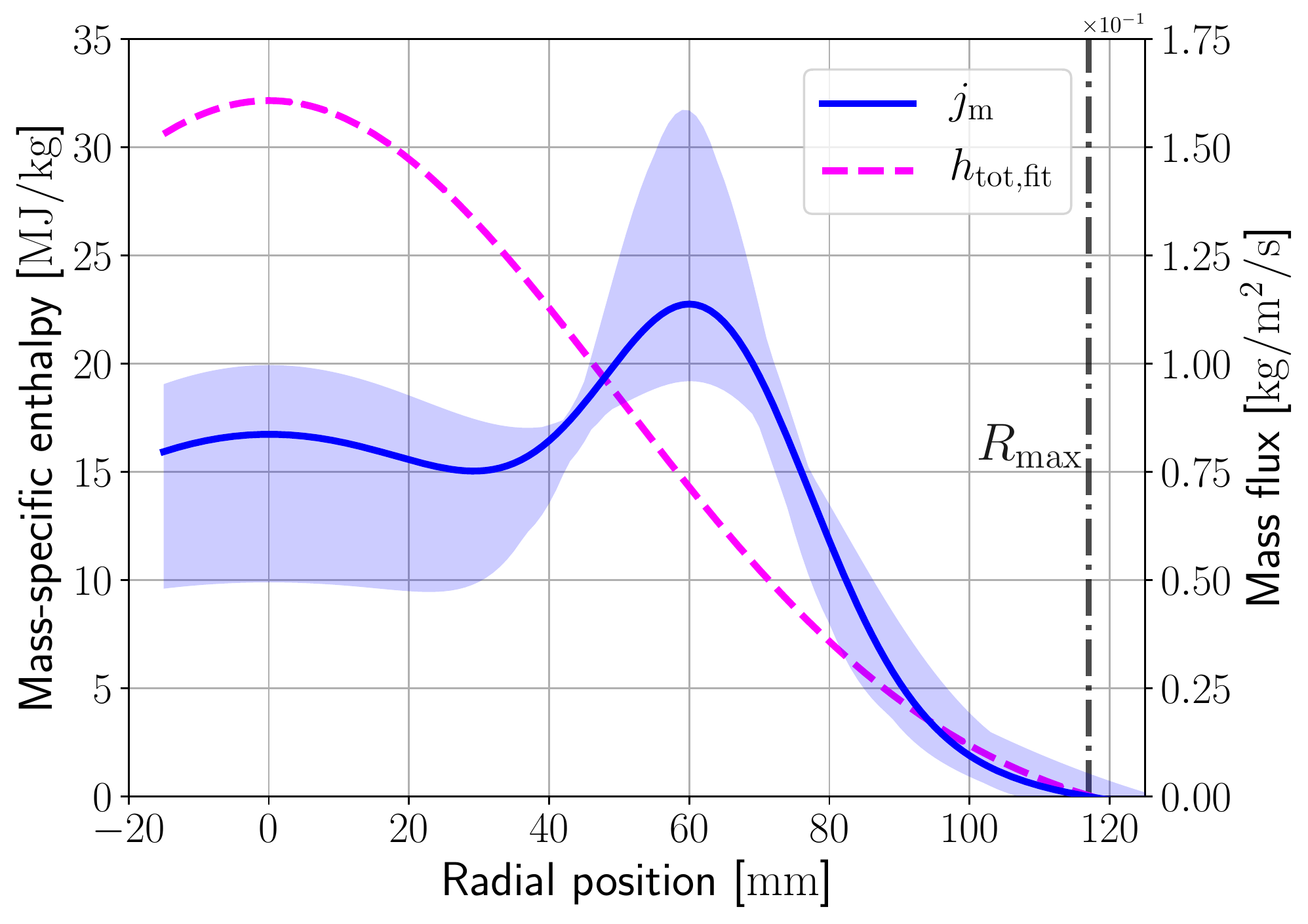}
	\caption{Reconstructed mass flux profile (blue line) for the condition CO2\#01b at $x=156\,\mathrm{mm}$ under the assumption of similarity to the Pitot profile. The Gaussian fit to the enthalpy data is plotted in magenta and the identified plasma edge is printed as a gray vertical line.}
	\label{fig:mass_flux}
\end{figure}

With the assumption of rotational symmetry, the Eqs. \ref{eq:mflux_boundary}-\ref{eq:power_integral} and the probe  measurement data, a mass flux profile with three degrees of freedom (DOF), such as simplified Gaussian profiles, can be clearly identified. However, for centered and off-centered Gauss profiles no solution of the mass flux can be found. Thus, it is further assumed that the shape of the radial mass flux distribution is similar to the Pitot pressure profile (Fig. \ref{fig:double_probe_hf_pitot}), since stagnation pressure and mass flux are closely related to one another. Consequently, the mass flux profile is believed to follow a double Gaussian distribution (Eq. \ref{eq:double_gauss}). The ratio of the profile widths $c_2/c_1$, as well as the height ratio of the two peaks (value of central peak divided by value of off-centered peak) are adopted from the Pitot profile to be 0.3 and 0.735, respectively. Under these conditions, a mass flux distribution can be clearly identified. For CO2\#01b the resulting profile is plotted in Fig. \ref{fig:mass_flux}.

At $x=156\,\mathrm{mm}$ a plasma edge position of $R_\mathrm{max}=117\,\mathrm{mm}$ was determined by examining the double probe results. This value is indicated in the figure. The calculated mass flux profile is plotted as a blue line. The profile follows a double Gaussian distribution with $a_1=8.92\cdot10^{-2}\,\mathrm{kg/m^2/s}$, $b_1=0\,\mathrm{m}$, $c_1=4.94\cdot10^{-2}\,\mathrm{m}$, $a_2=7.82\cdot10^{-2}\,\mathrm{kg/m^2/s}$, $b_2=6.30\cdot10^{-2}\,\mathrm{m}$, $c_2=1.48\cdot10^{-2}\,\mathrm{m}$ and $d=-5.50\,\mathrm{kg/m^2/s}$ (cp. Eq. \ref{eq:double_gauss}). The Gaussian fit of the enthalpy measurement is added in magenta, shifted horizontally to be axially symmetric. The light blue shadow serves as a sensitivity analysis towards the assumed values of $c_2/c_1$, the peak ratio and the plasma radius $R_\mathrm{max}$. The colored region includes all solutions of the mass flux profile for $0.15\leq c_2/c_1\leq 0.45$, $0.32 \leq $ peak ratio $ \leq 1.0$ and $107\,\mathrm{mm} \leq R_\mathrm{max} \leq 127\,\mathrm{mm}$. While in theory an infinite number of double Gaussian mass flux profiles can be found, most of them represent non-physical or unrealistic solutions, like a singularity at the position of mean enthalpy or a significantly higher mass flux on the centerline than at the off-centered peak position. Thus, it is probable that the sensitivity analysis covers a wide range of possible mass flux profiles. Nevertheless, other shapes than double Gaussian functions, e.g., with more degrees of freedom, could lead to valid solutions, but are not included in this analysis. It has to be noted that the global mass flux peak is close to the position of the mean value of the mass-specific enthalpy for CO2\#01b, which supports the choice of the profile shape. This might be due to a superposition of the tangential gas injection in IPG4 and the power coupling close the quartz tube wall due to the skin effect \cite{Herdrich2004}.\par
 
\subsubsection{CO\textsubscript{2} splitting under inhomogeneity}
In a last step, the mass flux distribution profile and the measured local mass-specific enthalpy values can be combined to calculate the integral CO\textsubscript{2} conversion and energy efficiency. A comparison to the splitting performance in the case of a homogeneous distribution of enthalpy over plasma radius quantifies the influence of the inhomogeneity. To determine the inhomogeneous splitting performance, the converted mass in the inhomogeneous case, assuming rotational symmetry, is calculated:
\begin{equation}
	\label{eq:conversion_inhom}
	\chi_\mathrm{inhom} = \frac{2 \pi}{\dot{m}} \int_{0}^{Rmax} \chi(r) j_\mathrm{m}(r) r \,\mathrm{d} r
\end{equation}
which leads to the energy efficiency in the inhomogeneous case:
\begin{equation}
	\label{eq:eta_inhom}
	\eta_\mathrm{inhom} =  \chi_\mathrm{inhom} \cdot \frac{\Delta H^\mathrm{0}_\mathrm{298}}{\bar{h}_\mathrm{tot}} \frac{e \cdot N_\mathrm{A}}{M_\mathrm{CO2}}
\end{equation}

\begin{figure}[hbt!]
	\centering
	\includegraphics[width=.49\textwidth]{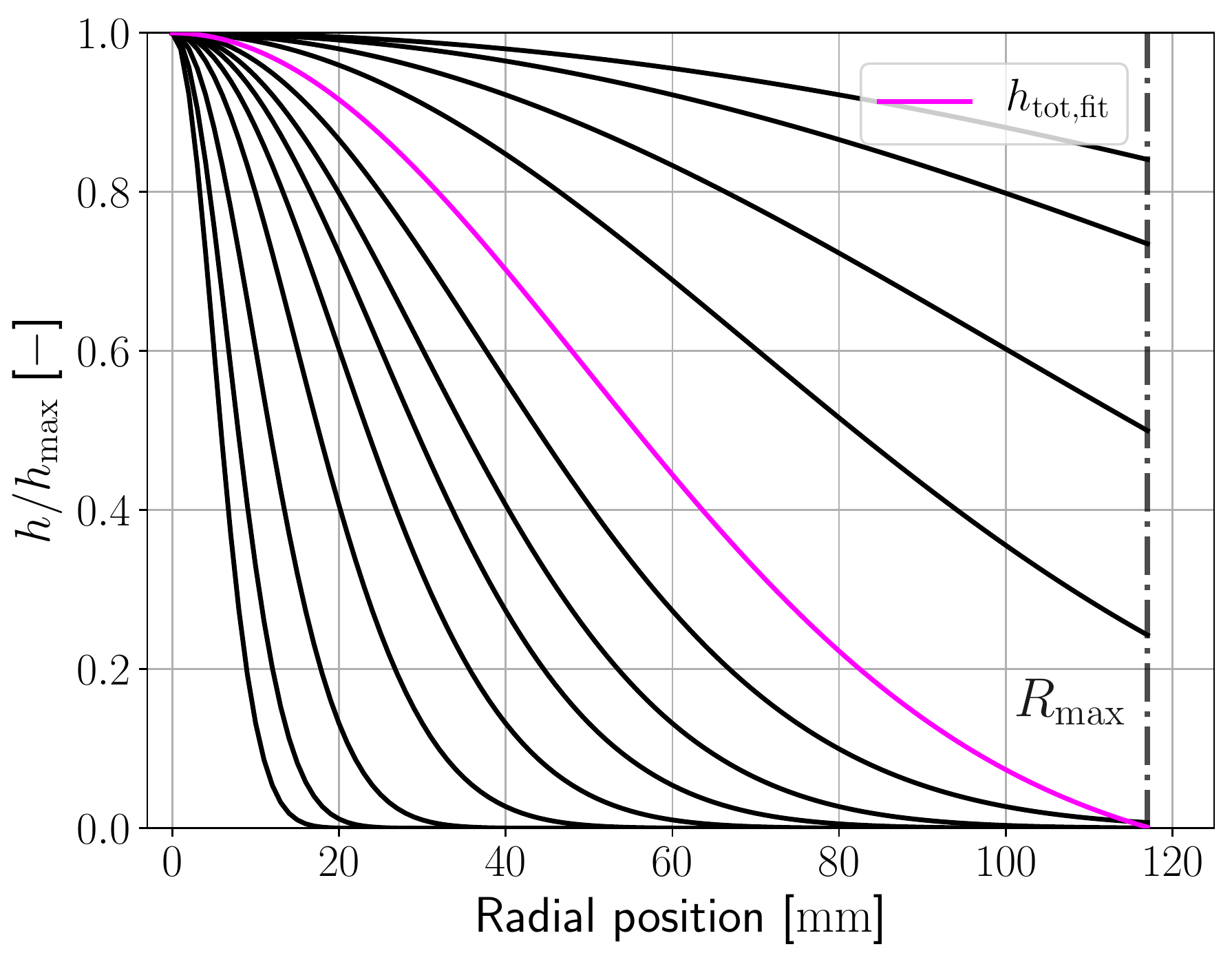}
	\caption{Normalized Gaussian radial enthalpy profiles with varying $\mathit{FWHM}$ used in the parameter study. The measured specific enthalpy distribution for CO2\#01b is colored in magenta.}
	\label{fig:enthalpy_profiles}
\end{figure}

The local CO\textsubscript{2} conversion values $\chi(r)$ are determined by feeding the local mass-specific enthalpy $h_\mathrm{tot,fit}(r)$ into the thermodynamic equilibrium simulation tool based on Cantera (cp. Section \ref{sec:CO2_splitting_cantera}). This methodology is used in a parameter study, where not only the measured enthalpy profile of CO2\#01b $h_\mathrm{tot,fit}$, but additional artificial Gaussian enthalpy distributions, varying in their Full Width at Half Maximum ($\mathit{FWHM}$), i.e., their spatial inhomogeneity, are investigated. The smaller the $\mathit{FWHM}$ of the enthalpy distribution, the higher is the inhomogeneity. The normalized enthalpy profiles used in the parameter study are plotted in Fig. \ref{fig:enthalpy_profiles}.

\begin{figure}[hbt!]
	\centering
	\includegraphics[width=.49\textwidth]{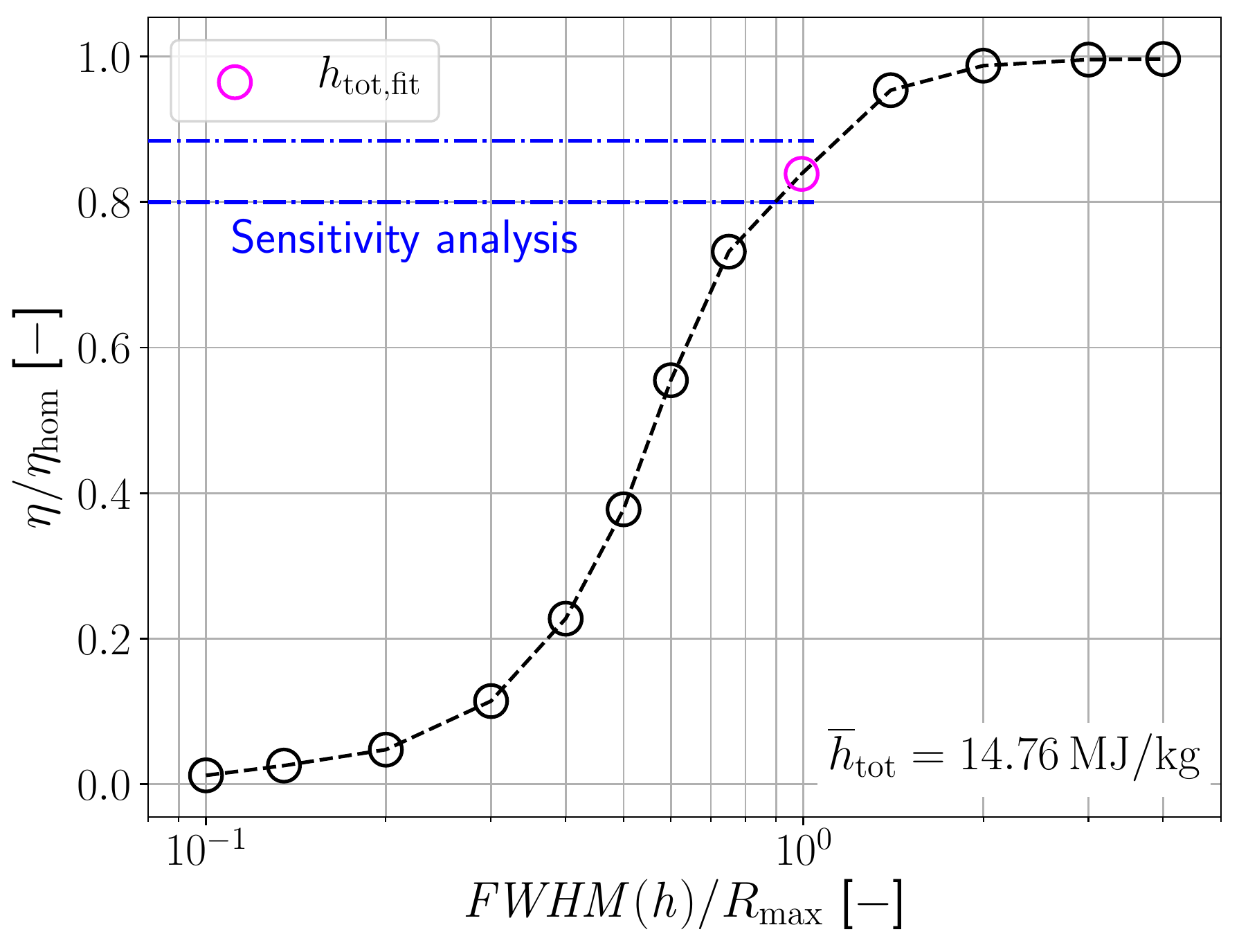}
	\caption{Ratio of inhomogeneous energy efficiency to the homogeneous comparison case as a function of the $\mathit{FWHM}$ of the mass-specific enthalpy profile, normalized with $R_\mathrm{max}$. Towards the left, the inhomogeneity increases. The ratio of the measured enthalpy distribution (CO2\#01b) is colored in magenta. Two horizontal blue lines represent the parameter range of the sensitivity analysis.}
	\label{fig:study_results}
\end{figure}

For each profile, the peak height and the $\mathit{FWHM}$ are chosen so that the bulk enthalpy equals $\bar{h}_\mathrm{tot}=14.76\,\mathrm{MJ/kg}$, which is the calorimetrically measured value for CO2\#01b. Moreover, the mass flux distribution is kept constant and equals the earlier determined profile (Fig. \ref{fig:mass_flux}) for the study. This theoretical parameter study would in reality resemble an experiment, where the mass injection into the generator remains unchanged, while the radial distribution (not the total value) of the applied power is altered. Consequently, the homogeneous comparison case for CO\textsubscript{2} splitting is the same for each distribution. The conversion and energy efficiency in the homogeneous case can be directly extracted from Fig. \ref{fig:cantera_splitting} to be $\chi_\mathrm{hom}=97.1\,\%$ and $\eta_\mathrm{hom}=42.2\,\%$, respectively. For the inhomogeneous enthalpy distributions, Eqs. \ref{eq:conversion_inhom} and \ref{eq:eta_inhom} are applied to calculate the CO\textsubscript{2} splitting performance in the case of spatial inhomogeneity. The results of the parameter study are plotted in Fig. \ref{fig:study_results}.

In the figure, the ratio of energy efficiency in the case of inhomogeneous specific energy distribution versus the homogeneous case is plotted as a function of the $\mathit{FWHM}$ of the enthalpy distribution, normalized with $R_\mathrm{max}$, which is a direct measure of inhomogeneity. It has to be emphasized again that all investigated profiles, in combination with the mass flux distribution, have the same mean enthalpy. Above all, the analysis shows that spatial inhomogeneity in specific enthalpy significantly lowers the CO\textsubscript{2} splitting performance in a plasma jet. This is due to the nonlinear correlation of SEI and the carbon dioxide conversion (cp. Fig. \ref{fig:cantera_splitting}). Moreover, it can be seen that the higher the inhomogeneity, the lower the splitting performance for the same mean mass-specific enthalpy, i.e., the same SEI. At large $\mathit{FWHM}$, the Gaussian specific enthalpy profile approaches a constant radial distribution, which results in an asymptotic behavior towards the homogeneous case. For CO2\#01b, the $\mathit{FWHM}$ is in the order of the plasma jet radius. The corresponding conversion and energy efficiency are $\chi_\mathrm{01b}=81.5\,\%$ and $\eta_\mathrm{01b}=35.4\,\%$, respectively. As a result, the inhomogeneous distribution of mass-specific enthalpy in IPG4 lowers the carbon dioxide splitting performance by $16\,\%$.\par
The blue horizontal lines in Fig. \ref{fig:study_results} indicate the sensitivity of the CO2\#01b splitting performance to changes in the mass flux profile. The plotted result space corresponds to the blue-shaded region in Fig. \ref{fig:mass_flux}. Although the mass flux profile is altered strongly in the sensitivity analysis, the influence on the splitting performance is quite limited (approx. $10\,\%$). The reason for this is that a large fraction of the mass flux is situated in outer regions of the plasma jet, where the mass flux variation is comparably small (cp. Fig. \ref{fig:mass_flux}). Moreover, the mean enthalpy in the study of $\bar{h}_\mathrm{tot}=14.76\,\mathrm{MJ/kg}$ is in a near linear region regarding the energy efficiency (cp. Fig. \ref{fig:cantera_splitting}). This region is characterized by robustness to inhomogeneities in the mass-specific enthalpy.\par

The inhomogeneities encountered in IPG4 are caused by the superposition of gas injection near the quartz tube edge and the radially inhomogeneous power distribution due to the skin effect \cite{Herdrich2002}. Therefore, the decrease in splitting performance is mostly defined by geometric parameters, like the injector placement, and the operational frequency. Since all conditions introduced in this paper (CO2\#01-CO2\#04) use the same plasma generator, operated at the same frequency, it is expected that the inhomogeneities are similar. Consequently, by changing the operating parameters the overall CO\textsubscript{2} splitting performance can be optimized, i.e., by adjusting the mean mass-specific enthalpy, but the inhomogeneity, leading to lower efficiencies, is not believed to be affected strongly. It has to be stated that this is only true as long as the discharge mode (inductive 'mode 3', cp. \cite{Georg2021power}) remains unchanged.

\section{Conclusion and outlook}
\label{sec:conclusion}
In this work, the specific energy inhomogeneity in the high-power carbon dioxide plasma jet of the inductive plasma generator IPG4 was determined experimentally. Its influence on the CO\textsubscript{2} splitting performance was quantified, showing a negative correlation between specific enthalpy inhomogeneity and CO\textsubscript{2} conversion, and thus energy efficiency.\par

Using a cavity calorimeter, calorimetric CO\textsubscript{2} plasma powers were measured for five different operating conditions in plasma wind tunnel PWK3. Moreover, total generator efficiencies of IPG4 in the range of $19.3-23.4\,\%$ were determined. A heat flux-Pitot double probe was applied to resolve the radial distribution of the mass-specific enthalpy for the condition CO2\#01b ($P_\mathrm{A}=160\,\mathrm{kW}$, $\dot{m}_\mathrm{CO2}=2.2\,\mathrm{g/s}$). The measurements showed a strong inhomogeneity in specific energy at the test position of $x=156\,\mathrm{mm}$. \par 
Based on the probe measurements, a radial mass flux profile was calculated. Here, a double Gaussian distribution, similar to the Pitot pressure, was assumed. 
In a parameter study, the influence of inhomogeneity in the mass-specific enthalpy distribution on the carbon dioxide splitting performance in thermodynamic equilibrium was quantified. 
The analysis revealed that specific energy inhomogeneities lower the CO\textsubscript{2} conversion, and consequently the energy efficiency, significantly, compared to a fully homogeneous distribution over plasma radius. Although this work only considers thermal CO\textsubscript{2} splitting, it is expected that the presented results qualitatively apply for non-thermal carbon dioxide dissociation as well, as long as some form of nonlinearity between SEI and the CO\textsubscript{2} conversion occurs.\par 
The shown influence of inhomogeneity on plasma-based CO\textsubscript{2} splitting should be considered in future designs of plasma sources, but also for the application of diagnostics and in plasma modeling, because splitting performance can be, and is, lost due to inhomogeneities. As a consequence, only local measurements are able to reveal the true potential of a plasma source with regards to plasma-based CO\textsubscript{2} conversion.\par
In future work, the determination of the local species composition in the plasma jet in PWK3 by optical emission spectroscopy (OES) is planned. This is to support or disprove the assumption of thermal CO\textsubscript{2} conversion in the plasma generator IPG4.

\section*{Funding}
This work was supported by the German Academic Scholarship Foundation (Studienstiftung des Deutschen Volkes); and the Friedrich und Elisabeth Boysen-Stiftung.









\end{document}